\documentclass[acmsmall]{acmart}

\usepackage{lipsum}

\usepackage[utf8]{inputenc}
\usepackage[USenglish]{babel}
\usepackage{colorprofiles}
\usepackage[a-2b,mathxmp]{pdfx}[2018/12/22]
\hypersetup{pdfstartview=}

\usepackage{float}
\usepackage{subfig}

\usepackage{pifont}

\usepackage{graphicx}
\graphicspath{ 
  {./fig/example/} 
  {./fig/illustration/} 
  {./fig/exp/} 
}

\usepackage{xcolor}
\definecolor{leafi-yellow}{RGB}{250, 231, 163}
\definecolor{leafi-green}{RGB}{202, 224, 184}
\definecolor{leafi-grey}{RGB}{208, 206, 206}
\definecolor{leafi-blue}{RGB}{193, 215, 236}
\definecolor{leafi-orange}{RGB}{240, 204, 177}
\definecolor{hercules-grey}{RGB}{217, 217, 217}

\usepackage{tikz}
\usetikzlibrary{tikzmark}
\usetikzlibrary{shapes.geometric}

\usepackage{enumitem}

\usepackage{bm}

\usepackage{url}

\usepackage{siunitx}

\usepackage{mathtools}

\usepackage{bm}

\usepackage{algorithm}
\usepackage{algorithmicx}
\usepackage[noend]{algpseudocode}

\algrenewcommand\algorithmicrequire{\textbf{Input:}}
\algrenewcommand\algorithmicensure{\textbf{Output:}}
\algnewcommand\algorithmiccontinue{\textbf{continue}}
\newcommand\Continue{\State \algorithmiccontinue}

\AtBeginDocument{%
  }

\setcopyright{authorsopen}

\acmJournal{PACMMOD}
\acmYear{2025} \acmVolume{3} \acmNumber{N1 (SIGMOD)}
\acmArticle{51} \acmMonth{2} \acmPrice{15.00}
\acmDOI{10.1145/3709701}

\received{July 2024}
\received[revised]{September 2024}
\received[accepted]{November 2024}

\begin{document}

\title{LeaFi: Data Series Indexes on Steroids with Learned Filters}

\author{Qitong Wang}
\affiliation{%
  \institution{LIPADE, Universit{\'e} Paris Cit{\'e}}
  \city{Paris}
  \country{France}
}
\email{qitong.wang@u-paris.fr}

\author{Ioana Ileana}
\affiliation{%
  \institution{LIPADE, Universit{\'e} Paris Cit{\'e}}
  \city{Paris}
  \country{France}
}
\email{ioana.ileana@parisdescartes.fr}

\author{Themis Palpanas}
\affiliation{%
  \institution{LIPADE, Universit{\'e} Paris Cit{\'e} \& French University Institute (IUF)}
  \city{Paris}
  \country{France}
}
\email{themis@mi.parisdescartes.fr}

\renewcommand{\shortauthors}{Qitong Wang, Ioana Ileana, \& Themis Palpanas}

\begin{abstract}
The ever-growing collections of data series create a pressing need for efficient similarity search, which serves as the backbone for various analytics pipelines. 
Recent studies have shown that tree-based series indexes excel in many scenarios. %
However, we observe a significant waste of effort during search, due to suboptimal pruning. 
To address this issue, we introduce LeaFi, a novel framework that uses machine learning models to boost pruning effectiveness of tree-based data series indexes. 
These models act as learned filters, which predict tight node-wise distance lower bounds that are used to make pruning decisions, thus, improving pruning effectiveness. 
We describe the LeaFi-enhanced index building algorithm, which selects leaf nodes and generates training data to insert and train machine learning models, as well as the LeaFi-enhanced search algorithm, which calibrates learned filters at query time to support the user-defined quality target of each query.
Our experimental evaluation, using two different tree-based series indexes and five diverse datasets, demonstrates the advantages of the proposed approach.
LeaFi-enhanced data-series indexes improve pruning ratio by up to 20x %
and search time by up to 32x, while maintaining a target recall of 99\%. 
\end{abstract}



\maketitle

\section{Introduction}
\label{sec-intro}

\begin{figure}[tb]
  \centering
  \subfloat[
      Average recall-at-1 compared to the number of searched DSTree leaf nodes.
      The search algorithm continues after finding the nearest neighbors.
    \label{fig:intro-motiv-num-nodes}]{
    \includegraphics[width=0.6\linewidth]{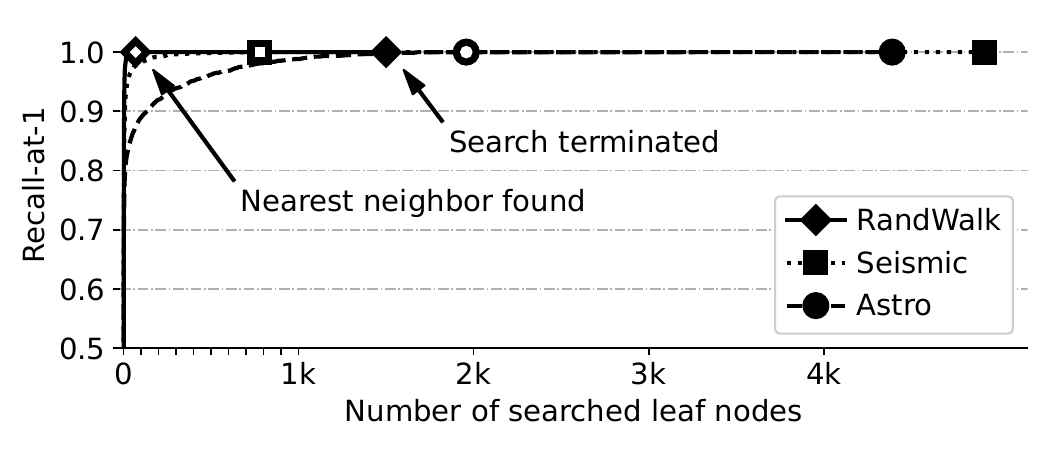}}\\
    
  \subfloat[
      Using the LeaFi predictions ($\star$) instead of current lower bounds ($\bm{+}$) boosts the pruning ratio for an Astro query from 23\% to 89.4\% (examples above the best-so-far curve can be pruned).
      Particularly, the pruning enhancement brought by LeaFi is observed both before and after the nearest neighbors are found.
      The LeaFi filters are trained to predict the optimal lower bounds ($\bm{\times}$), which can prune 99\% leaf nodes. 
    \label{fig:intro-motiv-node-nn}]{
    \includegraphics[width=0.6\linewidth]{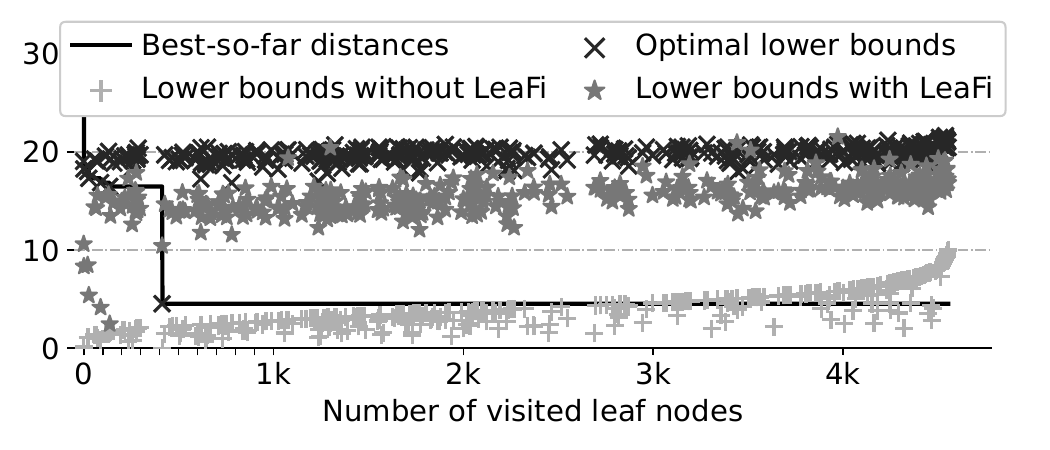}}\\
    
  \caption{
    A waste of data series search time, caused by insufficient pruning, is observed across various datasets.
    Employing the LeaFi predictions for the optimal lower bounds, instead of the current summarization-based lower bounds, improves the pruning ratios significantly.
  }
  \label{fig:intro-motiv-example}
\end{figure}

\noindent \textbf{\textit{Background.}}
With the rapid advancements and implementations of modern sensors, there is a significant rise in the generation, collection, and analysis of large datasets consisting of data series across various scientific fields~\cite{DBLP:journals/sigmod/PalpanasB19}.
Common techniques employed in the analysis of data series include classification~\cite{DBLP:journals/datamine/FawazFWIM19}, clustering~\cite{DBLP:conf/sigmod/PaparrizosG15}, pattern mining~\cite{DBLP:conf/kdd/ChiuKL03}, anomaly detection~\cite{DBLP:journals/pvldb/PaparrizosKBTPF22}, visualization~\cite{DBLP:journals/tvcg/GogolouTPB19}, etc.
However, as the data series collection grows largely in scale, it becomes crucial to incorporate \emph{similarity search} methods to maintain their efficiency and effectiveness~\cite{DBLP:journals/sigmod/Palpanas15}.
Data series similarity search is the technique used to identify the most similar series (usually the nearest neighbors) in a dataset, given a query series and a specific similarity measure (usually Euclidean distance~\cite{DBLP:journals/pvldb/DingTSWK08}).

Indexes are widely employed to speed up series similarity search.
Among various series indexing techniques, tree-based indexes have demonstrated state-of-the-art (SOTA) performance in numerous scenarios~\cite{DBLP:journals/pvldb/EchihabiZPB18, DBLP:journals/pvldb/EchihabiZPB19}, including in the context of hybrid solutions~\cite{DBLP:journals/pvldb/WeiPLP24,iliassigmod25,DBLP:journals/debu/00070P023}.
A tree-based index is built using lower-dimensional summarizations of data series~\cite{DBLP:conf/kdd/ShiehK08, DBLP:journals/pvldb/WangWPWH13}. 
Its index structure consists of internal nodes, which determine the order of visiting nodes, and leaf nodes, which store the original series values. 
Each node uses an aggregated summarized representation of the series it contains, in order to calculate a distance lower bound, which is compared with the best-so-far result to determine if this node can be pruned during query answering.

\noindent \textbf{\textit{Motivation.}}
Tree-based indexes greatly enhance the efficiency of series similarity search.
However, we argue that there are still significant opportunities for further acceleration.
This potential is highlighted in Figure~\ref{fig:intro-motiv-num-nodes} by the large gap in search time\footnote{
We use the number of searched leaf nodes as a hardware-agnostic surrogate for search time~\cite{DBLP:journals/pvldb/AziziEP23}.
In our context, \emph{searching} a leaf node means calculating the distances for all series it contains.
\emph{Visiting} a node, on the other hand, refers to checking its lower bound to determine whether it can be pruned.
A leaf node may be visited during a query, but not searched.
}, between when the nearest neighbor results are found and when they are actually returned~\cite{DBLP:conf/sigmod/GogolouTEBP20, DBLP:conf/sigmod/LiZAH20}. 
The search algorithm spends this significant amount of extra time trying to verify that there is no other better answer.
We attribute this inefficiency to the inability of existing tree-based series indexes to provide tight distance lower bounds for effective pruning. 
Figure~\ref{fig:intro-motiv-node-nn} reveals that only 23\% of the summarization-based lower bounds (cf. Figure~\ref{fig:intro-motiv-node-nn}, Lower bounds without LeaFi) exceed the best-so-far distances (cf. Figure~\ref{fig:intro-motiv-node-nn}, black line), resulting in a 23\% pruning ratio for a query on the Astro dataset, which is a rather poor performance.

Our key insight to this problem focuses on the introduction of the optimal lower bound for a leaf node, 
which is the smallest distance between a query and all the series that the leaf node contains; we call this distance the \emph{node-wise nearest neighbor distance}.
By leveraging these node-wise distances, 99\% of the leaf nodes (up from 23\%) can be pruned %
(cf. Figure~\ref{fig:intro-motiv-node-nn}, Optimal lower bounds).
However, directly calculating these distances by searching each leaf node is impractical. 
To overcome this challenge, we opt to employ machine learning models to predict the node-wise nearest neighbor distances. 
These models act as Learned Filters (LeaFi), improving the pruning effectiveness of tree-based series indexes.

\noindent \textbf{\textit{Our solution:} LeaFi.} 
In this paper, we introduce LeaFi, a novel general framework that introduces machine learning models into tree-based series indexes to improve their pruning capability for search acceleration.
We name these machine learning models \emph{learned filters}, and the index that uses them \emph{LeaFi-enhanced index}. 
In essence, LeaFi carefully places learned filters in a selected subset of leaf nodes, with each filter dedicated to one leaf node. 
These learned filters are trained to predict the node-wise nearest neighbor distances for a given query, which serve as lower bounds for comparison against the best-so-far distance.
Since the predictions are much tighter than the original node summarization-based lower bounds, the leaf node pruning ratios are largely improved. 
In our example, we go from 23\% pruning to 90\% pruning (cf. Figure~\ref{fig:intro-motiv-node-nn}, Lower bounds with LeaFi).
Though, this comes at the expense of a slight reduction in accuracy, which is controlled by the user and can be determined at query time (independently for each query). 
Note that a LeaFi-enhanced index can always provide exact results (guaranteed 100\% recall) for a specific query, simply by disabling the filter-based pruning strategy at query time. 
Our experimental evaluation shows that LeaFi-enhanced indexes achieve a remarkable improvement in pruning ratio (up to 20x more) %
and search time (up to 32x faster), while maintaining almost perfect accuracy (i.e., 99\% recall).

\begin{figure}[tb]
  \centering
  \includegraphics[width=0.6\linewidth]{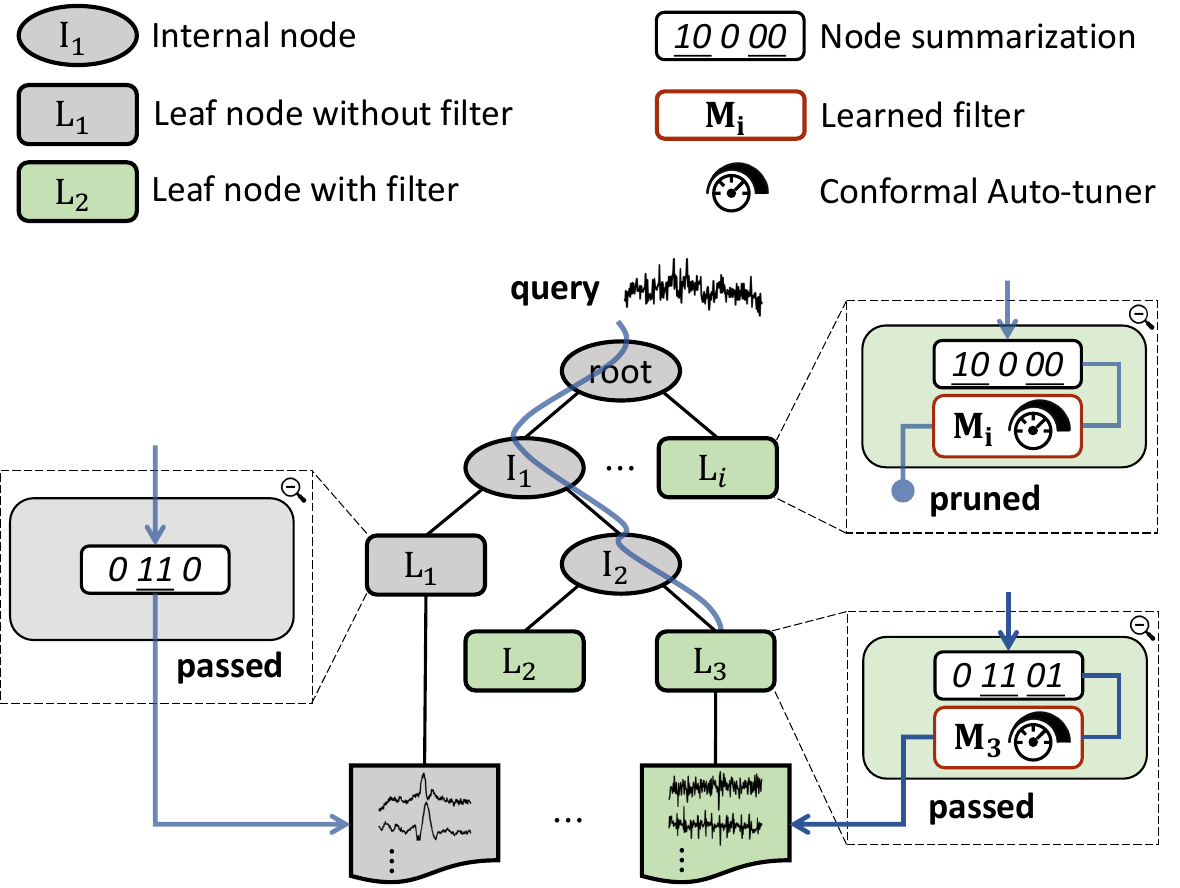}
  \caption{
    An illustration of a LeaFi-enhanced tree-based index structure, along with an example of its search procedure.
  }
  \label{fig:intro-index}
\end{figure}

Figure~\ref{fig:intro-index} illustrates the structure of a LeaFi-enhanced index.
For the selected leaf nodes $L_2$, $L_3$ and $L_i$, learned filters are paired with their node summarizations.
When these nodes are visited by a query, a cascade of summarization-based lower bounds and filter predictions are computed to determine whether these nodes should be pruned.
To the best of our knowledge, LeaFi is the first framework that incorporates machine learning models into data series indexes for improving pruning during similarity search.

\noindent \textbf{\textit{Technical challenges.}}
Incorporating learned filters into tree-based series indexes is not straightforward.
The LeaFi workflow unfolds in three main stages. %
It starts by selecting the subset of leaf nodes for filter insertion.
Next, it prepares the training data needed to train these filters.
The last step is to calibrate the filters' predictions to better serve as lower bounds.
We outline these steps along with their challenges and our solutions as follows.
 
First, effectively inserting learned filters requires identifying the subset of leaf nodes that offer the most significant reduction in search time. 
As modern machine learning models are accelerated using Graphics Processing Devices (GPU)~\cite{web/nvidia/cuda}, indiscriminately adding one learned filter to every leaf node could overwhelm the GPU memory.
Typical number of leaf node in tree-based series indexes can reach the range of 100K~\cite{DBLP:journals/pacmmod/Wang0WP023}).
Moreover, applying learned filters might not always improve search time.
For example, directly searching a small leaf node of 100 series can be faster than predicting a lower bound using a Multilayer Perceptron (MLP)~\cite{DBLP:journals/tc/Amari67} model. 
To address these issues systematically, we design a general formalization that treats leaf node selection as a knapsack problem~\cite{DBLP:books/daglib/0010031}. 
In this analogy, adding a learned filter to a leaf node is considered adding an "item", where its "value" is the expected reduction of this node's search time.
Given the constraint of available GPU memory, our goal in this setup is to identify which leaf nodes will yield the highest expected search time savings, when equipped with learned filters.
We further show that under certain assumptions, this general formalization can be simplified and solved by a greedy algorithm.

Second, training the inserted filters requires generating appropriate training data.
A key challenge is that the majority of the node-wise nearest neighbor distances fall out of the value range of global nearest neighbor distances.
Figure~\ref{fig:intro-motiv-node-nn}, shows that node-wise nearest neighbor distances lie around the value of 20, while the global nearest neighbor distance is 4.9.
Traditional training data generation methods, which typically involve random sampling combined with Gaussian noise~\cite{DBLP:journals/pvldb/EchihabiFZPB22}, tend to bias training towards these larger, node-wise distances, neglecting the global distances~\cite{DBLP:journals/pvldb/0003WNP22}. 
To resolve this issue, we propose a novel twofold strategy for training data generation. 
It consists of generating global training queries that are derived from the entire series collection and applicable to all leaf nodes, alongside node-wise training queries that are derived from and specific to each leaf node. 
Our twofold approach ensures that the training data encompasses both local (node-wise) and global contexts, facilitating unbiased filter training across all necessary value ranges.

Lastly, learned filters, being machine learning models, cannot ensure consistent prediction quality~\cite{GC:lecun2015deep}.
This compromises the exactness of the search result in original series indexes~\cite{DBLP:conf/kdd/WangP21}.
To mitigate the uncertainty of the search quality, we propose \emph{conformal auto-tuners}, a method inspired by conformal regressions~\cite{DBLP:journals/ftml/AngelopoulosB23} to enable support for user-requested search quality (e.g., recall, tightness, etc.) targets.
Our conformal auto-tuners calibrate the filter prediction using a learned offset, determined by a certain quality target that can be set at query time, independently for each query.
We employ a separate calibration set~\cite{DBLP:conf/ecml/PapadopoulosPVG02} to simulate search under different offsets and collect the result qualities.
The conformal auto-tuners learn the mapping between observed result qualities and corresponding calibration offsets, such that the calibration offset can be dynamically obtained for any given quality target.

\noindent \textbf{\textit{Contributions.}}
Our contributions can be summarized as follows.
\begin{enumerate}[noitemsep, topsep=0pt, wide=\parindent]
  \item We introduce the first approach that uses learned filters to improve the pruning effectiveness of tree-based series indexes, and hence accelerate data series similarity search.
  
  \item We design LeaFi, a novel general framework that effectively integrates learned filters into (different) tree-based series indexes. 
  LeaFi carefully selects an optimal subset of leaf nodes for filter insertion, and  generates appropriate training data for filter training.

  \item We propose conformal auto-tuners to mitigate the uncertainty in the results of machine learning models.
  Conformal auto-tuners calibrate learned filters at query time to support the user-defined quality target, independently for each query (the user may also choose to disable the learned filters).
  
  \item Our experimental evaluation, using two diverse tree-based series indexes and five diverse datasets, demonstrates the benefits of the proposed approach, and its advantages when compared to alternatives. %
  LeaFi-enhanced series indexes improve pruning ratio by up to 20x %
  and search time by up to 32x, while maintaining a target recall of 99\%. 
  Codes and datasets are available online\footnote{\url{https://github.com/qtwang/LeaFi}\label{fn:leafi-url}}.
\end{enumerate}

\newcounter{leafi-url}
\setcounter{leafi-url}{\value{footnote}}

\section{Related Work}
\label{sec-related-work}

\noindent \textbf{\textit{Data series indexes.}}
\label{sec-lit-index}
The most prominent data series indexing techniques can be categorized into graph-based indexes~\cite{DBLP:journals/pami/MalkovY20}, 
quantization~\cite{DBLP:journals/pami/GeHK014} and inverted indexes~\cite{DBLP:journals/pami/BabenkoL15}, 
locality-sensitive hashes~\cite{DBLP:journals/pvldb/HuangFZFN15},
and tree-based indexes~\cite{DBLP:conf/kdd/ShiehK08,DBLP:journals/pvldb/WangWPWH13}. 
Recent studies~\cite{DBLP:journals/pvldb/EchihabiZPB18, DBLP:journals/pvldb/EchihabiZPB19} have demonstrated that tree-based indexes~\cite{DBLP:conf/icde/PengFP20} achieve SOTA performance under several conditions (e.g., large-scale dataset).

iSAX~\cite{DBLP:conf/kdd/ShiehK08} and DSTree~\cite{DBLP:journals/pvldb/WangWPWH13} are two SOTA tree-based indexes for series similarity search of different strengths~\cite{DBLP:journals/pvldb/EchihabiZPB18, DBLP:journals/pvldb/EchihabiZPB19}.
iSAX is based on Symbolic aggregate approximation (SAX)~\cite{DBLP:conf/kdd/ShiehK08}, a discretized series summarization based on piecewise aggregate approximation (PAA)~\cite{DBLP:conf/sigmod/KeoghCMP01}.
PAA first transforms the data series into $l$ real values, and then SAX quantizes each PAA value using discrete symbols. 
iSAX (indexable SAX)~\cite{DBLP:conf/kdd/ShiehK08} enables the comparison of SAXs of different cardinalities, that makes SAX indexable through a prefix trie~\cite{DBLP:conf/aieeire/Briandais59}. 
DSTree~\cite{DBLP:journals/pvldb/WangWPWH13} is a dynamic splitting tree based on the adaptive piecewise constant approximation (EAPCA).
Furthermore, 
ADS+~\cite{DBLP:conf/sigmod/ZoumpatianosIP14} makes iSAX continuously adaptive to queries, 
ULISSE\cite{DBLP:journals/pvldb/LinardiP18,DBLP:journals/vldb/LinardiP20} supports variable-length queries, 
Coconut~\cite{DBLP:journals/pvldb/KondylakisDZP18} delivers a sortable iSAX variant, 
DPiSAX~\cite{DBLP:conf/icdm/YagoubiAMP17} and Odyssey~\cite{DBLP:journals/pvldb/ChatzakisFKPP23} make iSAX distributed, 
ParIS~\cite{DBLP:journals/tkde/PengFP21}, MESSI~\cite{DBLP:conf/icde/PengFP20,messijournal} and SING~\cite{DBLP:conf/icde/PengFP21} bring in modern hardware, FreSh~\cite{DBLP:conf/srds/FatourouKPP23} adds lock-freedom, 
Dumpy~\cite{DBLP:journals/pacmmod/Wang0WP023} and DumpyOS~\cite{DBLP:journals/vldb/WangWWPW24} introduce a data-adaptive multi-ary structure, 
while Hercules~\cite{DBLP:journals/pvldb/EchihabiFZPB22} and Elpis~\cite{DBLP:journals/pvldb/AziziEP23} combine the iSAX and EAPCA~\cite{DBLP:journals/pvldb/WangWPWH13} summarizations.

The proposed LeaFi framework is index agnostic.
We instantiate and evaluate it on both MESSI~\cite{messijournal} and DSTree~\cite{DBLP:journals/pvldb/WangWPWH13}, making its improvements translatable to most tree-based indexes.

\noindent \textbf{\textit{Machine learning applications in series indexes.}}
\label{sec-lit-learned-index}
Machine learning techniques have proven effective in enhancing various components of databases~\cite{DBLP:conf/sigmod/0001ZC21, DBLP:conf/sigmod/SaxenaRCLCCMKPN23}, such as indexes~\cite{DBLP:conf/sigmod/KraskaBCDP18, DBLP:conf/sigmod/NathanDAK20, DBLP:journals/pvldb/DingNAK20, DBLP:conf/icde/LiZSWT020, DBLP:conf/kdd/WangP21}, cardinality estimators~\cite{DBLP:conf/sigmod/Sun0021, DBLP:conf/sigmod/KimJSHCC22, DBLP:journals/pacmmod/WuNAKM23}, etc.
A few existing works are also motivated by the fact that there is waste in search time~\cite{DBLP:conf/sigmod/GogolouTEBP20, DBLP:journals/vldb/EchihabiTGBP23}. 
These works can be divided into two categories, early stopping approaches~\cite{DBLP:journals/pvldb/EchihabiZPB19, DBLP:conf/sigmod/GogolouTEBP20, DBLP:journals/vldb/EchihabiTGBP23} and leaf node reordering approaches~\cite{SC:kang2021case}.

In the context of data series similarity search, $\epsilon$-search identifies heuristic stopping criteria when best-so-far results are in the $\epsilon$ neighborhood of nearest neighbor results~\cite{DBLP:journals/pvldb/EchihabiZPB19}.
$\delta\epsilon$-search extends $\epsilon$-search by supporting a confidence level $\delta$, based on estimated pairwise distance distribution~\cite{DBLP:journals/pvldb/EchihabiZPB19}.
Progressive Search (ProS) incorporates machine learning models to estimate when the exact results are retrieved, using the query and best-so-far distances~\cite{DBLP:conf/sigmod/GogolouTEBP20, DBLP:journals/vldb/EchihabiTGBP23}.
Learned Reordering (LR) determines the visiting order of the leaf nodes by predicting their probabilities of containing the nearest neighbor results for a given query~\cite{SC:kang2021case}.

\begin{figure}[tb]
  \centering
  
  \subfloat{
  \includegraphics[width=0.66\linewidth]{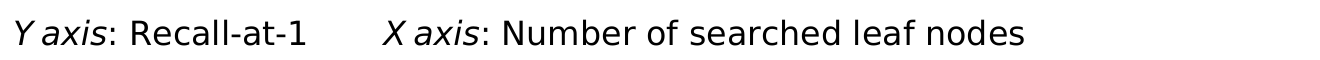}}\\[-2ex]
  
  \setcounter{subfigure}{0}
  \subfloat[Early stopping
    \label{fig:liter-optimal-early-stopping}]{
    \includegraphics[width=.22\linewidth]{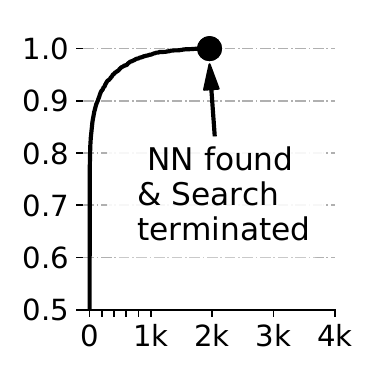}}
  \subfloat[Reordering
    \label{fig:liter-optimal-reordering}]{
    \includegraphics[width=.22\linewidth]{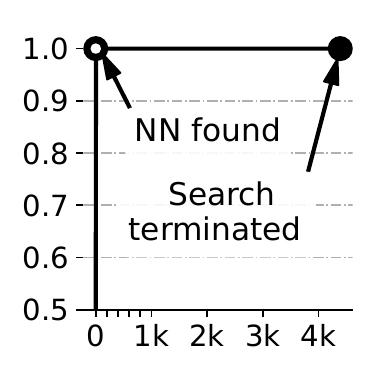}}
  \subfloat[LeaFi
    \label{fig:liter-optimal-leafi}]{
    \includegraphics[width=.22\linewidth]{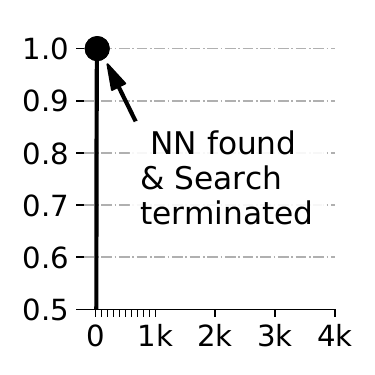}}
    
  \vspace*{-0.2cm}
  \caption{The optimal search time that can be possibly achieved by early-stopping approaches~\cite{DBLP:journals/pvldb/EchihabiZPB19, DBLP:conf/sigmod/GogolouTEBP20, DBLP:journals/vldb/EchihabiTGBP23}, leaf node reordering approaches~\cite{SC:kang2021case} and LeaFi for DSTree index on Astro dataset.
  The axes are the same as Figure~\ref{fig:intro-motiv-node-nn} (x-axis in Figure~\ref{fig:liter-optimal-leafi} has a different scale).
  }
  \label{fig:liter-comparison}
\end{figure}

Figure~\ref{fig:liter-comparison} illustrates the improvement potential for DSTree index on Astro dataset, as in Figure~\ref{fig:intro-motiv-num-nodes}, of early stopping approaches ($\epsilon$-search, $\delta\epsilon$-search, ProS and FLT), leaf node reordering approaches (LR), and our proposed learned filter approach (LeaFi).
These optimal performance is simulated by assuming all machine learning models make no mistake.
Leaf node reordering approaches help series indexes find the nearest neighbor results in the first node being visited and employ the tightest best-so-far distance.
However, as shown in Figure~\ref{fig:intro-motiv-node-nn}, a tight best-so-far distance provide marginal help in pruning more leaf nodes - most of the summarization-based lower bounds are still smaller than that. 
Early stopping approaches can terminate search right after the nearest neighbor results are retrieved, but it cannot reduce the search time before the retrieval. 

On the other hand, LeaFi helps series indexes to search only those leaf nodes that can update the best-so-far distances, as the tightest lower bounds are employed in pruning decision, hence attaining a significant and consistent improvement potential.
Moreover, as trained using the node-wise distance information instead of index-wise leaf node searching information, LeaFi can be efficiently trained using 2k examples (0.002\% of the collection), compared to 100k to 1m examples (0.1\% of the collection) in the literature~\cite{DBLP:conf/sigmod/LiZAH20, SC:kang2021case}.
As far as we are aware, LeaFi represents the first framework that incorporates learned filters in data series indexes and provides substantial improvement in pruning ratio and search time.

FAISS Learned Termination (FLT) is an early-termination technique proposed for vector similarity search using kNN graphs~\cite{DBLP:conf/sigmod/LiZAH20}.
Its stopping criterion is predicted by a nontrivial expert-crafted feature set, which cannot be directly applied to tree-based series indexes, and, in contrast to LeaFi, it does not offer a mechanism to set search quality targets.

\noindent \textbf{\textit{Conformal regressions.}}
\label{sec-lit-conformal}
Conformal regression is a statistical approach that enhances existing regression methods by providing predictive intervals with a guarantee on their coverage probability~\cite{GS:vovk2005algorithmic}. 
It involves fitting a regression model to a dataset and then generating predictions that include an interval which, with a specified level of confidence (e.g., 95\%), is expected to cover the true target values. 
This process relies on the computation of nonconformity scores that measure how unusual new observations are compared to training data~\cite{DBLP:conf/ecml/PapadopoulosPVG02, DBLP:journals/ftml/AngelopoulosB23}.

However, the direct application of existing conformal regression techniques cannot help LeaFi support the user-requested search quality.
This is because the conformal prediction intervals are derived independently for each model, whereas achieving an expected target of a LeaFi-enhanced search result, e.g., 99\% recall, requires tuning all learned filters collaboratively at the same time.
Hence, in LeaFi, we further design our auto-tuning approach based on the conformal regression framework.
To the best of our knowledge, LeaFi demonstrates the first effort to introduce conformal regression techniques into the domain of learned indexes.

\section{Preliminaries}
\label{sec-background}

\noindent \textbf{\textit{Data series.}}
\label{sec-bkg-series}
A \emph{data series}, $S=\{p_1, ..., p_m\}$, is a sequence of points, where each point $p_i=(v_i,t_i)$, $1 \le i \le m$ is associated to a real value $v_i$ and a position $t_i$. 
We call $m$ the \emph{length} of the series. 
$\mathcal{S}$ denotes a collection of data series, i.e., $\mathcal{S}=\{S_1, ..., S_n\}$. 
We call $n$ the \emph{size} of the series collection.

A \emph{summarization} $E=\{e_1, ..., e_l\}$ of a series $S$ is its lower-dimensional representation, that preserves some desired properties of $\mathcal{S}$.
For example, SAX in MESSI~\cite{DBLP:conf/kdd/ShiehK08} and APCA in DSTree~\cite{DBLP:journals/pvldb/WangWPWH13} are two popular series summarizations.
In series similarity search, summarizations can be utilized to calculate lower bounds between a series or a set of series and a query.

\noindent \textbf{\textit{Similarity search.}}
\label{sec-bkg-search}
Given a query series $S_q$, a series collection $\mathcal{S}$ of size $n$, a distance measure $d$, \emph{similarity search} targets to identify the series $S_c \in \mathcal{S}$ whose distance to $S_q$ is the smallest, i.e., $\forall S_o \in \mathcal{S}, S_o \neq S_c, d(S_c, S_q) \leq d(S_o, S_q)$.

The LeaFi framework works for any distance measure supported by the backbone index, including Euclidean and Dynamic Time Warping (DTW), two popular distances for series similarity search~\cite{DBLP:journals/pvldb/DingTSWK08}.

\noindent \textbf{\textit{Tree-based indexes.}}
\label{sec-bkg-index}
Tree-based series indexes, including DSTree~\cite{DBLP:journals/pvldb/WangWPWH13} and MESSI~\cite{DBLP:conf/kdd/ShiehK08}, are constituted by \emph{internal nodes} $I_i$s and \emph{leaf nodes} $L_i$s, as shown in Figure~\ref{fig:intro-index}.
We use $N_i$ to denote a node when there is no need to distinguish between it being an internal node or leaf node.
Only leaf nodes store the raw series.
An internal node routes a series to one of its child node that this series should be inserted into.
Both types of nodes contain a \emph{node summarization} $E^{N}_{i}$ that aggregates the series summarizations of all series it contains.
Node summarizations are used to calculate distance lower bounds for search routing and leaf node pruning.

During query answering, internal nodes navigate the query series $S_q$ to visit %
leaf nodes %
according to their lower bounds to the query.
Only the leaf nodes whose subtree cannot be pruned are visited. %
We maintain a best-so-far result $d^{\text{bsf}}(S_q,N_i)$, i.e., the smallest distance before visiting a node $N_i$.
Given a node $N_i$, we first calculate the node summarization-based lower bound $d^{\text{lb}}(S_q, N_i)$ for the distances between $S_q$ and all series $N_i$ contains.
When the context is clear, we remove inputs $(\cdot)$ or subscripts $_{i}$ in the equations for clarity.
We then compare $d^{\text{lb}}$ with $d^{\text{bsf}}$.
$d^{\text{lb}}>d^{\text{bsf}}$ indicates all series in %
a leaf node $L_i$ or a subtree $I_i$%
have larger distances to $S_q$, thus %
$L_i$ or $I_i$%
can be safely pruned.
Otherwise, there might be a series that has smaller distance to $S_q$. %
For $L_{i}$, %
we have to search $L_{i}$ to get its node-wise nearest neighbor distance $d^{\text{L}}(S_{q}, L_{i})= \min_{k} d(S_{q}, S_{k}), S_{k} \in L_{i}$, and check whether $d^{\text{L}}$ can update $d^{\text{bsf}}$.
For $I_{i}$, we repeat the pruning checking for the nodes in its subtree. %
After all nodes are either visited or pruned, %
$d^{\text{bsf}}$ is returned as the true nearest neighbor result.

\noindent \textbf{\textit{Conformal regression.}}
\label{sec-bkg-conformal}
Conformal regression utilizes posterior statistics to auto-tune machine learning predictions to target at a certain confidence level~\cite{DBLP:journals/ftml/AngelopoulosB23}.
Suppose we collect a training set $\{(x_1, y_1), (x_2, y_2), ..., (x_l, y_l)\}$ of $l$ examples.
In our case, $x_i$ is a training query (i.e., $S_q$) and $y_i$ is the node-wise nearest neighbor distance $d^{\text{N}}$.
Given a confidence level $1-\delta$, \emph{conformal regression}~\cite{DBLP:journals/ftml/AngelopoulosB23} predicts an interval, instead of a single value, such that the true $y_{l+1}$ of a new example $x_{l+1}$ will be covered by this interval with a $1-\delta$ confidence.

In LeaFi, we embrace the core designs from the classic \emph{inductive conformal regression}~\cite{DBLP:conf/ecml/PapadopoulosPVG02} as a proof-of-concept study.
Inductive conformal regression splits the training data into two subsets: the \emph{proper training set} $\{(x_1, y_1), (x_2, y_2), ..., (x_m, y_m)\}$ and the \emph{calibration set} $\{(x_{m+1}, y_{m+1}), (x_{m+2}, y_{m+2}), \\..., (x_l, y_l)\}$.
After training a machine learning model on the proper training set, we calculate and sort the \emph{non-conformity measures} on the calibration set: $\alpha_{i} \coloneqq |y_{m+1}-\hat{y}_{m+i}|, i=1, 2, ..., l-m$, where $\hat{y}_{m+i}$ is the prediction of a sample $x_{m+i}$.
We use $[\alpha_{(1)}, \alpha_{(2)}, ..., \alpha_{(l-m)}]$ to denote the descending-ordered $\alpha$s on the calibration set.
Let $j_s=|\{\alpha_{i}: \alpha_{i} \geq \alpha_{(s)}\}|, s=1, 2, ..., l-m$ be the number of $\alpha$s that are at least as large as $\alpha_{(s)}$.
Given the confidence level $1-\delta$ and a new example $x_{l+1}$, the \emph{predictive region} is derived as $(\hat{y}_{l+1}-\alpha_{(s)}, \hat{y}_{l+1}+\alpha_{(s)})$, by choosing the $j_s$ such that $\delta=j_{s}/(l-m+1)$. 

In Section~\ref{sec-method-conformal-auto-tuning}, we further discuss how to establish the mapping between $\alpha$s and search result quality, and then identify the corresponding $\alpha$s to auto-tune the filter predictions for the support of search quality targets.

\section{The \texorpdfstring{L\MakeLowercase{ea}F\MakeLowercase{i}}{LeaFi} Framework}
\label{sec-method}

In this section, we present the algorithm details of enhancing data series indexes with learned filters.
We start with developing the notations for a LeaFi-enhanced index structure. 
Then, in Section~\ref{sec-method-workflow}, we outline the LeaFi-enhanced series index building and search workflow, illustrated in Figure~\ref{fig:meth-workflow}.
The details of leaf node selection, training data generation and filter conformal auto-tuning, are presented in Section~\ref{sec-method-node-selection}, Section~\ref{sec-method-training-data-generation} and Section~\ref{sec-method-conformal-auto-tuning}, respectively.

As illustrated in Figure~\ref{fig:intro-index}, a LeaFi-enhanced tree-based series index contains learned filters in a selected subset of its leaf nodes.
A learned filter $M_i$ in a node $L_i$ is a trained machine learning model, which takes in a query $S_q$ and predicts $d^{\text{L}}$.
We denote the prediction as $d^{\text{f}}(S_q, M_i)$, and the conformal adjusted prediction as $d^{\text{F}}(S_q, M_i)$.
Hence, the learned filter-based pruning decision can be made by comparing the best-so-far distance $d^{\text{bsf}}$ with the predicted lower bound $d^{\text{F}}$.
If $d^{\text{F}} > d^{\text{bsf}}$, we prune $L_i$, otherwise we search $L_i$.

\subsection{LeaFi-enhanced Series Index Workflow}
\label{sec-method-workflow}
\begin{figure}[tb]
  \centering

  \subfloat[The LeaFi index building workflow.
    \label{fig:meth-workflow-offline}]{
    \includegraphics[width=0.6\linewidth]{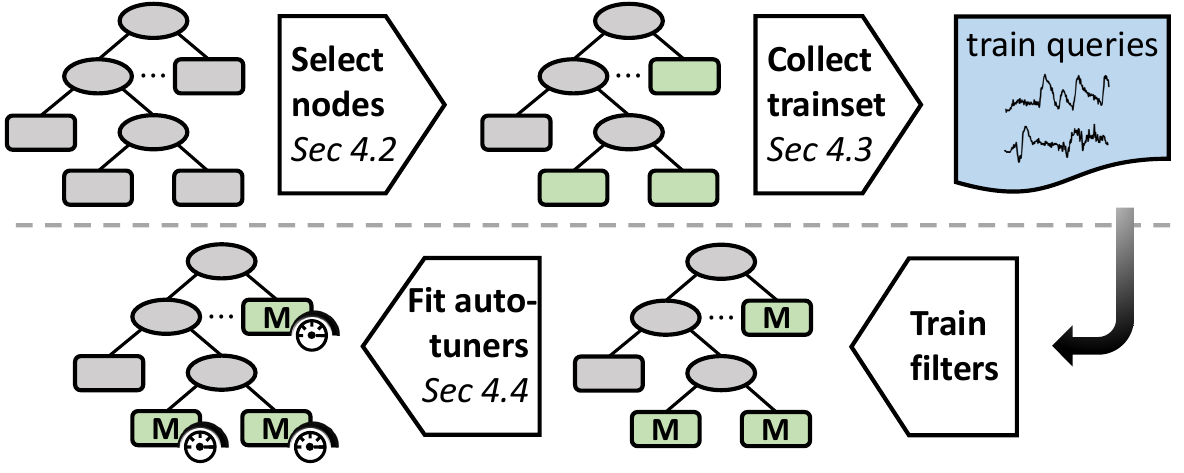}}\\ %
  
  \subfloat[The LeaFi query answering workflow.
    The quality target can be auto-tuned for each individual query.
    \label{fig:meth-workflow-online}]{
    \includegraphics[width=0.64\linewidth]{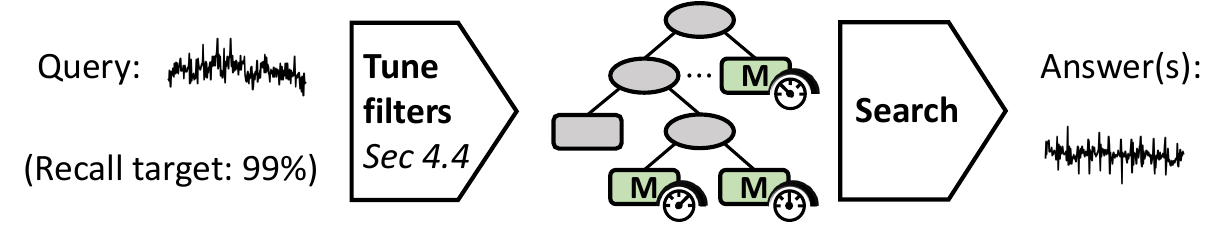}}
  
  \caption{
  The index building and query answering workflow of a LeaFi-enhanced series index.
  }
  \label{fig:meth-workflow}
\end{figure}

We modify both the index building workflow and the search workflow of original tree-based series indexes to enable learned filters, as illustrated in Figure~\ref{fig:meth-workflow}.
We outline these procedures as follows, and present the details in later sections.

\subsubsection{LeaFi-enhanced Index Building Overview}
\label{sec-method-index-building}
For index building, we add three main steps after the original index is built, illustrated in Figure~\ref{fig:meth-workflow-offline}.
These steps are (a) select leaf nodes for filter insertion, (b) generate training data from both global (index-wise) context and local (node-wise) context, and train the inserted filters, and (c) collect conformal training data and fit the conformal auto-tuners.

\noindent \textbf{\textit{Select leaf nodes (Section~\ref{sec-method-node-selection}).}}
We first establish a general framework that selects the optimal subset of leaf nodes that maximize the benefit of learned filter insertion, using knapsack solvers~\cite{DBLP:books/daglib/0010031}.
We show that, under certain assumptions, we can simplify the framework and select leaf nodes using a greedy algorithm.
In brief, LeaFi calculates a leaf node size threshold $th$, based on the number of distance calculations that execute in the same time as one filter inference.
We then select the leaf nodes according to their sizes until $th$ is met, or the GPU memory is fully occupied.

\noindent \textbf{\textit{Generate training data (Section~\ref{sec-method-training-data-generation}).}}
We generate training data $\mathcal{S}_{q}$ from both global and local contexts.
Index-wise training data $\mathcal{S}_{q(g)}$ is sampled from the whole dataset and shared among all selected leaf nodes, while node-wise training data $\mathcal{S}_{q(l)}$ is sampled individually from each node and is only available to their corresponding node. %
There is no additional expense by splitting $\mathcal{S}_{q}$ into $\mathcal{S}_{q(g)} \cup \mathcal{S}_{q(l)}$.
The collection time for $\mathcal{S}_{q(g)}$ and $\mathcal{S}_{q(l)}$ node-wise data is %
the same as for $n_{q}=n_{q(g)}+n_{q(l)}$ index-wise training data.

\noindent \textbf{\textit{Fit conformal auto-tuners (Section~\ref{sec-method-conformal-auto-tuning}).}}
After inserted filters fit the training set, we fit the conformal auto-tuners to enable the support for user-request search quality targets.
The absolute prediction errors $\{\alpha_{i}\}$, i.e., the non-conformity scores in the conformal regression context~\cite{DBLP:journals/ftml/AngelopoulosB23}, are employed as offsets to auto-tune the predictions. %
One auto-tuner is added to each learned filter.

\begin{algorithm}[tb]
  \caption{LeaFi-enhanced Index Building}
  \label{algo:inde-building}
  \begin{algorithmic}[1]
  
  \Require a series collection $\mathcal{S}$ of size $n$, a tree-based index $I$, the training data size $n_{q}=n_{q(g)}+n_{q(l)}$, and the available GPU memory $c^{\text{M}}$
  \Ensure the LeaFi-enhanced index $I^{\text{F}}$
  \State $\{N^{\text{F}}_{i}\}$ = \Call{SelectLeafNode}{$I$, $c^{\text{M}}$}
  \State $\mathcal{S}_{q(g)}$ = \Call{GenerateGlobalQueries}{$\mathcal{S}$}
  \State $T_{g} \coloneq \{(d^{bsf}, d^{\text{L}})\}$ = \Call{SearchGlobalQueries}{$I$, $n_{q(g)}$, $\{N^{\text{F}}_{i}\}$}
  \State $\mathcal{S}_{q(l)} = \emptyset$, $T_{l} = \emptyset$
  \ForAll{$N^{\text{F}}_{i} \in \{N^{\text{F}}_{i}\}$}
    \State $\mathcal{S}_{q(l), i}$ = \Call{GenerateLocalQueries}{$N^{\text{F}}_{i}$, $n_{q(l)}$, $T_{g}$}
    \State $T_{l,i} \coloneq \{d^{\text{L}}\}$ = \Call{SearchLocalQueries}{$N^{\text{F}}_{i}$, $n_{q(l)}$,}
    \State $\mathcal{S}_{q(l)} = \mathcal{S}_{q(l)} \cup \mathcal{S}_{q(g), i}$, $T_{l} = T_{l} \cup T_{l,i}$
  \EndFor
  \State $I^{\text{f}}$ = \Call{TrainFilters}{$\{N^{\text{F}}_{i}\}$, $\mathcal{S}_{q(g)}$, $T_{g}$, $\mathcal{S}_{q(l)}$, $T_{l}$}
  \State $I^{\text{F}}$ = \Call{FitAutoTuners}{$I^{\text{f}}$, $\mathcal{S}_{q(g)}$, $T_{g}$}
  \\ \Return $I^{\text{F}}$
  
  \end{algorithmic}
\end{algorithm}

\noindent \textbf{\textit{Index building pseudocode.}}
We describe the LeaFi-enhanced index building pseudocode for a general tree-based index building algorithm in Algorithm~\ref{algo:inde-building}.
The sub-functions S\textsc{elect}L\textsc{eaf}N\textsc{ode} is defined later in Algorithm~\ref{algo:node-selection}
and F\textsc{it}A\textsc{uto}T\textsc{uners} in Algorithm~\ref{algo:fit-auto-tuners}.

Given a series collection $\mathcal{S}$, a tree-based index $I$, the training data size $n_{q}=n_{q(g)}+n_{q(l)}$, and the available GPU memory $c^{\text{M}}$, we first select the subset of leaf nodes $\{N^{\text{F}}_{i}\}$.
We then generate the collect the index-wise training data from line 2 to line 3.
From line 5 to line 8, we iterate over each selected leaf node $N^{\text{F}}_{i}$ and generate its corresponding node-wise training data. 
Line 9 trains all the inserted filters using the both index-wise and node-wise training data.
We finalize the LeaFi-enhanced index building workflow by fitting the conformal auto-tuners in line 10.

\subsubsection{LeaFi-enhanced Search Overview}
\label{sec-method-search}
There are two new operations in LeaFi-enhanced search workflow, illustrated in Figure~\ref{fig:meth-workflow-online}, compared to the original search workflow.
Before searching for a query, we first auto-tune the filters according to the user-request search quality target.
Then, we visit each leaf node, and check whether it can be pruned using a cascade of summarization-based strategy and learned filter-based strategy as in Figure~\ref{fig:intro-index}.

\noindent \textbf{\textit{Auto-tune learned filters (Section~\ref{sec-method-conformal-auto-tuning}).}}
Given a user-specified search quality target $q_{j}$, we use the learned mapping $f^{c}$ to calculate the corresponding adjusting offset $o_{i,j}=f^{c}(q_{j})$ for each filter $M_{i}$.
Then, we use the left boundary $d^{f}_{i}-o_{i,j}$ of the adjusted interval as the predicted lower bound $d^{F}_{i,j}$, to be compared with $d^{bsf}$ for the pruning decision of node $N_{i}$.

\noindent \textbf{\textit{Search pseudocode.}}
We describe the LeaFi-enhanced search pseudocode for a general tree-based index search algorithm in Algorithm~\ref{algo:search}.

Given a query $S_{q}$, a LeaFi-enhanced index $I^{\text{F}}$, the result quality target $Q$, and the number of nearest neighbors $k$, we first auto-tune the learned filters based on the target $Q$.
A hash table is employed to avoid redundant computation for the same $Q$s, which we omitted in Algorithm~\ref{algo:search} for simplicity.
We then iterate across the leaf nodes according to the order suggested by $I^{\text{F}}$.
When visiting a node $N_{i}$, we first check whether it can be pruned using $d^{\text{lb}}$ in line 4.
Otherwise, we check whether $N_{i}$ can be pruned using $d^{\text{F}}$, if a learned filter $M_{i}$ resides in $N_{i}$ in line 6.
If neither $d^{\text{lb}}$ nor $d^{\text{F}}$ can prune $N_{i}$, we search $N_{i}$ for $S_{q}$ to obtain its node-wise nearest neighbor results ${\mathcal{S}_{r, i}}$ and update ${\mathcal{S}_{r}}$ from line 8 to line 10.

\subsection{Leaf Node Selection}
\label{sec-method-node-selection}

In this section, we present how the problem of leaf node selection can be formalized as a knapsack problem~\cite{DBLP:books/daglib/0010031} in Section~\ref{sec-method-knapsack}.
We show that this general formalization can be simplified under certain assumptions and solved efficiently by a greedy algorithm.

Briefly speaking, our greedy leaf node selection algorithm works as follows.
First, we sort all leaf nodes $\{N_{i}\}$ by their sizes $\{|N_{i}|\}$ in a non-increasing order.
Then, we select those nodes whose sizes exceed a threshold $th$ to insert filters, until all available GPU memory is occupied.
We describe how $th$ is derived from a simplified knapsack formalization and how it is determined in Section~\ref{sec-method-greedy-selection}.

\begin{algorithm}[tb]
  \caption{LeaFi-enhanced Search}
  \label{algo:search}
  \begin{algorithmic}[1]
  
  \Require a query $S_{q}$, a LeaFi-enhanced index $I^{\text{F}}$, the result quality target $Q$, and the number of nearest neighbors $k$
  \Ensure the search results ${\mathcal{S}_{r}}$
  \State $I^{\text{F}}$ = \Call{AutoTuneFilters}{$I^{\text{F}}$, $Q$}
  \State $d^{\text{bsf}}$ = INF, ${\mathcal{S}_{r}}$ = \Call{PriorityQueue}{$\emptyset$, $k$}
  \ForAll{$N_{i} \in [N_{j}]$, ordered by $I^{\text{F}}$}
    \If{$d^{\text{lb}}>d^{\text{bsf}}$} 
      \Continue
    \EndIf
    \If{$M_{i} \in N_{i}$ $\And$ $d^{\text{F}} > d^{\text{bsf}}$} 
      \Continue
    \EndIf
    \State $S_{r,i}$, $d^{\text{L}}$ = \Call{SearchNode}{$N_{i}$, $d^{\text{bsf}}$}
    \If{$d^{\text{L}} <= d^{\text{bsf}}$} 
      \State $d^{\text{bsf}} = d^{\text{L}}$, ${\mathcal{S}_{r}} = {\mathcal{S}_{r}} \oplus S_{r,i}$
    \EndIf
  \EndFor
  \\ \Return ${\mathcal{S}_{r}}$
  
  \end{algorithmic}
\end{algorithm}

\subsubsection{A general solution framework as a knapsack problem.}
\label{sec-method-knapsack}
We start with a brief review of the knapsack problem, and then built the analogy between the knapsack problem and leaf node selection.

The knapsack problem is a classic example of combinatorial optimization~\cite{DBLP:books/daglib/0010031}.
It is a type of resource allocation problem where the objective is to maximize the total value of items placed in a knapsack without exceeding its capacity.
Given a set of $n$ items, each with a weight $w_i$ and a value $v_i$, and a knapsack with a weight capacity $W$, the goal is to determine the subset of items to include in the knapsack such that the total weight does not exceed $W$ and the total value is maximized.
This can be represented by the following optimization problem: maximize $\sum^n_{i=1}v_i x_i$, subjective to $x_i\in\{0, 1\}$ and $\sum^n_{i=1}w_i x_i \leq W$.
Here, $x_i$ is a binary decision variable that indicates whether item $i$ is included ($x_i=1$) or excluded ($x_i=0$) in the knapsack.
There are several variants of the knapsack problem, among which we only consider 0/1 knapsack problem in this paper.

The analogy between a knapsack problem is built as the following.
For $n^N$ leaf nodes, we consider that there are $n^N$ filters, each of which is corresponding to one specific node.
In this case, a filter $M_i$ can be considered as an item.
Its value is the expected reduction of search time, denoted by $b_i$, induced by inserting $M_i$ into its corresponding leaf node $N_i$.
The weight of an item $M_i$ is its GPU memory footprint, and the weight capacity of the knapsack is the volume of available GPU memory $c^{\text{M}}$.
Hence, by solving the mapped knapsack problem, we identify the subset of leaf nodes that obtain the optimal search time improvement by learned filter insertion.
We formalize the leaf node selection as a knapsack problem in Equation~\ref{eq:bnf-index}:
\begin{equation}
\begin{aligned} 
\label{eq:bnf-index}
  \max \quad \sum_{i} &\; b_{i}x_{i} \\
  \textrm{s.t.} \quad \sum_{i} &\; w_{i}x_{i} \leq c^{\text{M}} \\
  &\; x_{i} \in \{0, 1\}, \forall i = 1, ..., n^{\text{N}}. 
\end{aligned}
\end{equation}
\noindent where $w_{i}$ and $c^{\text{M}}$ can be calculated analytically~\cite{DBLP:conf/sigsoft/GaoLZLZLY20} or measured empirically using GPU profiling tools~\cite{web/nvidia/nvml}.
In this work, we only consider the case where all learned filters share the same neural network architecture.
As a result, they also share the same memory footprint, i.e., $w_{i}=w, \forall i = 1, ..., n^{\text{N}}$.
This general framework can also tackle the case where different filters have different network architectures by extending it to a multiple-choice knapsack problem~\cite{DBLP:books/daglib/0010031}.
We will study this case in our future work.

\noindent \textbf{\textit{Formulate the expected time reduction $b_{i}$.}}
To solve Equation~\ref{eq:bnf-index}, we need to estimate the only unknown variable $b_{i}$ , i.e., the reduced search time by adding a filter $M_{i}$ to a specific leaf node $N_{i}$.
We observe that $b_{i}$ is influenced by the following factors: the node size $|N_{i}|$, the summarization-based pruning probability $p^{\text{lb}}$, the filter-based pruning probability $p^{\text{F}}$.
A larger $|N_{i}|$ or a smaller $p^{\text{lb}}$ indicates the original index $I$ spent more effort on node $N_{i}$, and a larger $p^{\text{F}}$ hints that filter $M_{i}$ works well for node $N_{i}$.
Hence, $b_{i}$ can be formulated using $|N_{i}|$, $p^{\text{lb}}$ and $p^{\text{F}}$ as Equation~\ref{eq:bnf-node}:
\begin{align}
\label{eq:bnf-node}
  b_{i} = (1-p^{\text{lb}}_{i}) \times (p^{\text{F}}_{i} \times t^{\text{S}} \times |N_{i}| - t^{\text{F}})
\end{align}
\noindent where $t^{\text{S}}$ denotes the distance calculation time for one series and $t^{\text{F}}_{j}$ denotes the inference time for filter $M_{i}$.
Note that $b_{i}$ can be negative in cases where $|N_{i}|$ is small or $t^{\text{F}}$ is large.
We describe how we prohibit such insertions by automatically establishing a threshold for $|N_{i}|$ using $t^{\text{S}}$ and $t^{\text{F}}$ in Section~\ref{sec-method-greedy-selection}.

\noindent \textbf{\textit{Challenge: estimate the filter-based pruning probability $p^{\text{F}}$.}}
Although we can measure the wall-clock time for $t^{\text{S}}$ and $t^{\text{F}}$ empirically using trial experiments,
there are two other variables in Equation~\ref{eq:bnf-node} that cannot be directly calculated or measured, i.e., the summarization-based pruning probability $p^{\text{lb}}$ and the filter-based pruning probability $p^{\text{F}}$.

In our preliminary studies, we find $p^{\text{lb}}$ can be estimated by collecting $d^{\text{lb}}$ along with $d^{\text{bsf}}$ and $d^{\text{L}}$ using trial experiments.
However, estimating $p^{\text{F}}$ means to accurately estimate the performance of machine learning models without fully fitting their respective training data. 
This is a nontrivial task, and has been under extensive studies in the automated machine learning (AutoML) literature~\cite{DBLP:conf/kdd/YangFWU20, DBLP:journals/csur/KarmakerHSXZV22}.
Considering the positioning of LeaFi is a proof-of-concept study to demonstrate the potential of embracing learned filters in series indexes, we opt to simplify Equation~\ref{eq:bnf-node} by assuming that $p^{\text{lb}}$ and $p^{\text{F}}$ are the same across leaf nodes.

\begin{algorithm}[tb]
  \caption{Leaf Node Selection}
  \label{algo:node-selection}
  \begin{algorithmic}[1]
  
  \Require the leaf nodes $\{N_{i}\}$ of index $I^{\text{F}}$, the available GPU memory $c^{\text{M}}$ and hyperparameter $a$
  \Ensure a selected subset of leaf nodes $\{N^{F}_{i}\}$ for filter insertion
  
  \State $w$ = \Call{MeasureFilterGPUMemory}{$M$}
  \State $t^{\text{F}}$ = \Call{MeasureFilterInferenceTime}{$M$}
  \State $t^{\text{S}}$ = \Call{MeasureDistanceCalculationTime}{$\cdot$}
  \State $th$ = $a\times t^{\text{F}} / t^{\text{S}}$
  \State $[N_{(i)}]$ = \Call{SortBySize}{$\{N_{i}\}$}
  \State ${\mathcal{N}_{r}}$ = $\emptyset$, $w^{\text{M}}$ = $0$
  \ForAll{$N_{(j)} \in [N_{(i)}]$}
    \If{$|N_{(j)}| \geq th$ $\And$ $w^{\text{M}} < c^{\text{M}}$} 
      \State ${\mathcal{N}_{r}}$ = ${\mathcal{N}_{r}} \oplus N_{(j)}$, $w^{\text{M}}=w^{\text{M}}+w$
    \EndIf
  \EndFor
  \\ \Return ${\mathcal{N}_{r}}$
  
  \end{algorithmic}
\end{algorithm}

\subsubsection{The Greedy Selection Algorithm.}
\label{sec-method-greedy-selection}
To tackle the challenge of accurately estimating the performance of machine learning models, we propose to assume that $p^{\text{lb}}$ and $p^{\text{F}}$ are the same across leaf nodes in LeaFi.
That is, $\forall i = 1, ..., n^{\text{N}}$, both $p^{\text{lb}}_{i}=p^{\text{lb}}$ and $p^{\text{F}}_{i}=p^{\text{F}}$ hold.
Although this assumption is coarse-grained, we show later in Equation~\ref{eq:bnf-threshold} that it can derive a safe greedy algorithm that has no negative effect.
Under this assumption, Equation~\ref{eq:bnf-node} can be simplified into Equation~\ref{eq:bnf-simplification}:
\begin{align}
\label{eq:bnf-simplification}
  b_{i} = (1-p^{\text{lb}}) \times (p^{\text{F}} \times t^{\text{S}} \times |N_{i}| - t^{\text{F}})
\end{align}
\noindent where $b_{i}$ is only correlated to $|N_{i}|$ ($t^{\text{S}}$ and $t^{\text{F}}$ are also the same across leaf nodes).
Moreover, the correlation between $b_{i}$ and $|N_{i}|$ is positive.
That is, a larger node size $|N_{i}|$ introduces larger search time reduction $b_{i}$ under this assumption.
Hence, we can sort the leaf nodes based on their sizes and select them the larger nodes until consuming all available GPU memory.

However, as mentioned in Section~\ref{sec-method-node-selection}, $b_{i}$ can be negative in cases where $|N_{i}|$ is small or $t^{\text{F}}$ is large, which should be prohibited.
To formalize the constraint, we set $b_{i}>0$, transform Equation~\ref{eq:bnf-simplification} and get a threshold $th$ for leaf node size $|N_{i}|$ in Equation~\ref{eq:bnf-threshold}:
\begin{align}
\label{eq:bnf-threshold}
  b_{i} > 0 \Rightarrow \; |N_{i}| > \frac{1}{p^{\text{F}}} \times \frac{t^{\text{F}}}{t^{\text{S}}} \Rightarrow \; |N_{i}| > a \frac{t^{\text{F}}}{t^{\text{S}}}
\end{align}
\noindent where $a \coloneq 1/p^{\text{F}}$ is a hyperparameter.
$t^{\text{F}} / t^{\text{S}}$ is the number of series that its distance calculation time is equivalent to the inference time of a learned filter.
By its definition $1/p^{\text{F}}$, $a$ has an intrinsic interpretation as being the inverse of the filter-based pruning probability $p^{\text{F}}$.
Hence, it is intuitive to set $a$ by choosing a lower bound for $p^{\text{F}}$ to ensure $b_{i} > 0$.
In our experiments, we set $a=2$ (i.e., $p^{\text{F}}=50\%$) and evaluated that it worked well across different index prototypes, datasets, query noise levels and search quality targets.

\noindent \textbf{\textit{Leaf node selection pseudocode.}}
Putting all these designs together, we present our leaf node selection approach of the LeaFi index building workflow in Algorithm~\ref{algo:node-selection}.

We first collect all necessary runtime statistics, including the GPU memory footprint of a filter in line 1, the filter inference time in line 2 and the series distance calculation time in line 3.
Then, we calculate the node size threshold $th$ based on Equation~\ref{eq:bnf-threshold} in line 4.
In line 5, we sort all the leaf node in a non-increasing order in line 5.
From line 6 to line 9, for each leaf node $N_{(j)}$, if its size $|N_{(j)}|$ exceeds $th$ and there is GPU memory available, we add $N_{(j)}$ into the set of selected leaf nodes ${\mathcal{N}_{r}}$.
After checking all leaf nodes, we return ${\mathcal{N}_{r}}$ as in line 10.

\subsection{Training Data Generation}
\label{sec-method-training-data-generation}
After inserting filters into the selected subset of leaf nodes, we need to train these filters using proper training data.

\noindent \textbf{\textit{Challenge: imbalanced target value range.}}
However, we observe that even in the presence of a large real workload, preparing training data for the filters is nontrivial.
The challenge lies in the fact that the majority of the node-wise nearest neighbor distances fall out of the value range of global nearest neighbor distances.
This observation holds in general, because the tree-based series indexes are expected to group similar series into the same leaf nodes.
Only a few leaf nodes should have similar series to the query, which have small node-wise nearest neighbor distances close to the global nearest neighbor distances.

\begin{figure}[tb]
  \centering
  
  \subfloat[
  Only global training data ($\bm{\times}$ in figure, $\mathbb{G}$ in table) for leaf node $L_1$.
  The Node column indicates in which node each series finds its nearest neighbor.
    \label{fig:method-global-examples-only}]{
    \includegraphics[width=.6\linewidth]{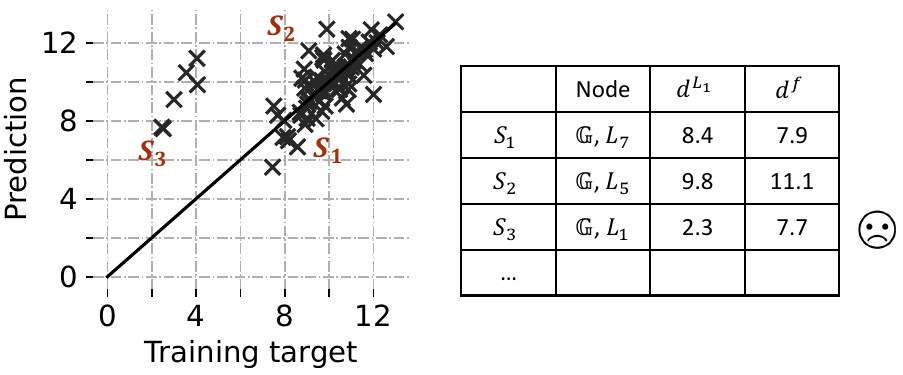}} \\
    
  \subfloat[
  Both global ($\bm{\times}$) and local ($\star$ in figure, $\mathbb{L}$ in table) training data. %
    \label{fig:method-global-and-local-examples}]{
    \includegraphics[width=.6\linewidth]{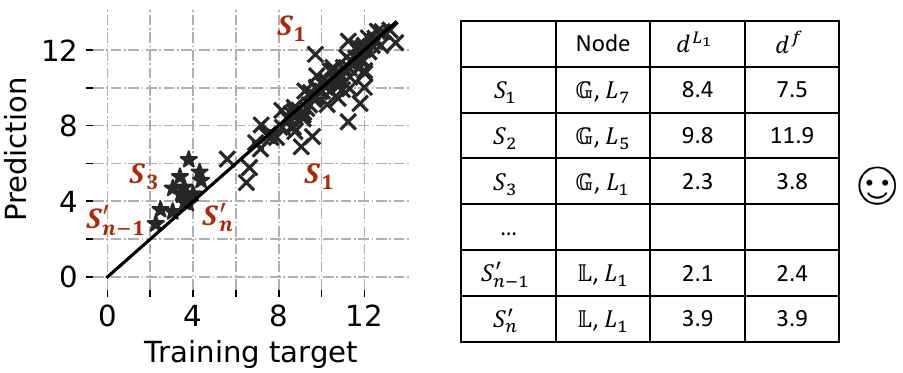}}
    
  \caption{
  The value distribution of the training targets (x-axis) for filters.
  Considering the global examples only overlooks the more important small value ranges, where the global nearest neighbor results are found.
  }
  \label{fig:method-motivate-local-examples}
\end{figure}

Figure~\ref{fig:method-motivate-local-examples} further illustrates this observation and demonstrates its pollution in filter training.
In Figure~\ref{fig:method-global-examples-only}, only five example fall into the global nearest neighbor distance range around $[3, 5]$, %
and only $S_3$ finds its nearest neighbor in $L_i$ with a distance 2.3. %
The majority of the node-wise nearest neighbor distances are located in $[7.5, 12.5]$. 
Thus, the model being trained on targets of $[7.5, 12.5]$, also predicts $d^f(S_3, L_1) = 7.7$, which is much larger than the actual $2.3$.
This causes the model to generate inaccurate predictions, hence, requiring large adjusting ranges, and leading to small pruning enhancements. %
This disparity not only brings harmful target bias on filter training~\cite{DBLP:journals/ml/RibeiroM20, DBLP:conf/icml/YangZCWK21}, but also invalidates the conformal regression as the absolute prediction errors will possess difference different distributions for the examples in different target ranges~\cite{DBLP:conf/ecml/PapadopoulosPVG02, GS:vovk2005algorithmic}.

\noindent \textbf{\textit{Our solution: generating both global and local data.}}
To resolve this issue, we propose a novel twofold strategy for training data generation. 
It consists of generating global training queries that are derived from the entire series collection and applicable to all leaf nodes, alongside node-wise training queries that are derived from and specific to each selected leaf node. 
As illustrated in Figure~\ref{fig:method-global-and-local-examples}, our twofold approach ensures that the training data encompasses both global (index-wise) and local (node-wise) contexts, providing unbiased training target value ranges.

In LeaFi, we employ the conventional query generating approach from the data series literature~\cite{DBLP:journals/pvldb/EchihabiZPB18, DBLP:journals/pvldb/EchihabiZPB19, DBLP:journals/pvldb/EchihabiFZPB22} for both index-wise and node-wise training data.
We uniformly select $n_{q(g)}$ random series $\mathcal{S}_{q(g)}$ from the entire collection, as well as $n_{q(l)}$ random series $\mathcal{S}_{q(l), i}$ from each select leaf node $N^F_{i}$, and then add random Gaussian noise.
We then search the index for $\mathcal{S}_{q(g)}$ to collect the training targets $T_{g} \coloneq \{(d^{bsf}, d^{\text{L}})\}$, i.e., the node-wise nearest neighbor distances $\{d^{\text{L}}\}$ and the best-so-far distances $\{d^{\text{bsf}}\}$.
$\{d^{\text{bsf}}\}$ is utilized later to fit the auto-tuners for the support of user-requested search quality targets.
For each set of node-wise training series $\mathcal{S}_{q(l), i}$, we only search its corresponding leaf node $N^F_{i}$ to collect the training targets $T_{l,i}\coloneq \{d^{\text{L}}\}$.
To train the filter $M_i$ inserted into $N^F_{i}$, we use $\mathcal{S}_{q(g)} \cup \mathcal{S}_{q(l), i}$ as the input and $T_{g} \cup T_{l,i}$ as the targets.
In our experiments, we empirically set the split $n_{q(g)}/n_{q(l)}=3$ to obtain a balanced target values ranges.
This process is presented in lines~5-8 of Algorithm~\ref{algo:inde-building}.

\begin{figure}[tb]
  \centering
  
  \includegraphics[width=0.5\linewidth]{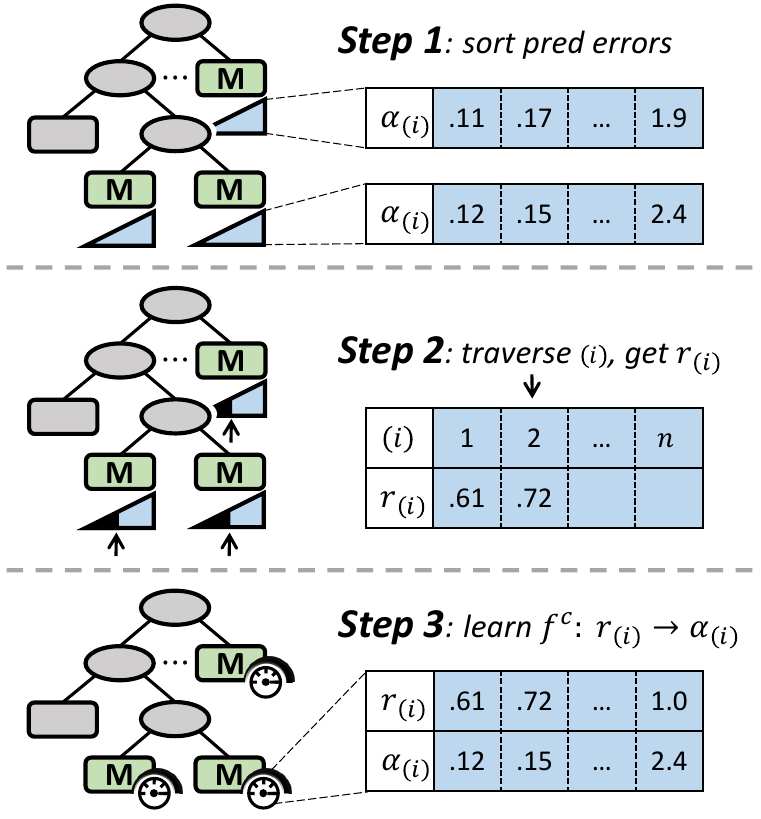}
  \caption{
  Collect training examples on calibration set for conformal auto-tuners, and learn the mapping between search result qualities and prediction adjusting offsets.
  }
  \label{fig:meth-fit-autotuner}
\end{figure}

\subsection{Filter Conformal Auto-tuning}
\label{sec-method-conformal-auto-tuning}
In this section, we describe how the LeaFi-enhanced indexes can be auto-tuned to support some user-defined qualities of the query answers.
The conformal auto-tuners make use of the prediction statistics from a separate calibration set~\cite{DBLP:conf/ecml/PapadopoulosPVG02} to build a mapping between the search quality and prediction adjusting offset for each trained filter.
In LeaFi, the calibration set is a subset of the index-wise training data.

Briefly speaking, we employ absolute prediction errors $\alpha$ as the adjusting offsets $o$, following the common exercise of inductive conformal regression~\cite{DBLP:conf/ecml/PapadopoulosPVG02}.
During LeaFi-enhanced index building, the training data for auto-tuners are the examples of the mapping between offsets $o$ and search result qualities $q$, simulated and collected using the calibration set.
We then fit a spline regression~\cite{steffen1990simple} on these examples to obtain a learned mapping $f^{c}: q \to o$.
In LeaFi-enhanced search, given a user-requested quality target $q$, we first determine the corresponding adjusting offset $o_i = f^{c}_i(q)$ for each learned filter $M_i$.
Then, the adjusted filter prediction $d^{\text{F}}_{i}=d^{\text{f}}_{i}-o_i$ is calculated to help prune node $N_i$ if necessary.

\subsubsection{Conformal Auto-tuner Learning in Index Building.}
\label{sec-conformal-index-building}
Figure~\ref{fig:meth-fit-autotuner} shows how we prepare the training data for the conformal auto-tuners in three steps.

First, we calculate the absolute prediction errors $\{\alpha\}_{i}$ on the calibration set for each learned filter $M_i$.
$\{\alpha\}_{i}$ are then employed as the candidate adjusting offsets ${o}_i$.
To avoid enumerating all possible combinations of candidate offsets for all inserted leaf nodes, we sort ${o}_i$ (Step 1 in Figure~\ref{fig:meth-fit-autotuner}), such that we can iterate over the sorted positions $(j)$, instead of enumerating the combinations. %
Second, we iterate over sorted positions $(j)$ to get the corresponding result quality $q_{(j)}$ on the calibration set.
For each $(j)$, we select the corresponding error $o_{i,(j)}$ as the adjusting offset of $M_i$.
We then simulate LeaFi-enhanced search for the calibration queries, and calculate the result quality $q_{(j)}$ (Step 2 in Figure~\ref{fig:meth-fit-autotuner}). %
After checking all sorted positions, we collect the examples $\{(q_{(j)}, o_{i,(j)}\}$ of the mapping between the result quality and the adjusting offset for each $M_i$.
Using these examples, we train a spline regression model~\cite{steffen1990simple} as the auto-tuner to fit the mapping $f^{c}_{i}: r \to \alpha$ (Step 3 in Figure~\ref{fig:meth-fit-autotuner}).%

\begin{algorithm}[tb]
  \caption{Auto-tuner Learning}
  \label{algo:fit-auto-tuners}
  \begin{algorithmic}[1]
  
  \Require a set of learned filters $\{M_{i}\}$ for index $I^{\text{F}}$, a set of calibration series $\mathcal{S}_{q(c)}$
  \Ensure learned auto-tuners $\{f^c_i\}$
  
  \ForAll{$M_{i} \in \{M_{i}\}$}
    \State $E_i$ = $\emptyset$
    \State $\{\alpha\}_i$ = \Call{CalAbsPredErrors}{$M_{i}$, $\mathcal{S}_{q(c)}$}
    \State $[\alpha]_i$ = \Call{Sort}{$\{\alpha\}_i$}
  \EndFor
  \ForAll{$j$ $\in$ $[1, |\mathcal{S}_{q(c)}|]$}
    \ForAll{$M_{i} \in \{M_{i}\}$}
      \State $o_{j, i}$ = $\alpha_{i, (j)}$
    \EndFor
    \ForAll{$S_{q(c), k} \in \mathcal{S}_{q(c)}$}
      \State $\mathcal{S}_{r, k}$ = \Call{SimulateSearch}{$I^{\text{F}}$, $\{o_{i}\}_{j}$, $S_{q(c), k}$}
      \State $q_{j, k}$ = \Call{EvaluateQuality}{$\mathcal{S}_{r, k}$}
      \ForAll{$M_{i} \in \{M_{i}\}$}
        \State $E_i$ = $E_i \oplus (q_{j, k}, o_{j, i})$
      \EndFor
    \EndFor
  \EndFor
  \ForAll{$M_{i} \in \{M_{i}\}$}
    \State $f^c_i$ = \Call{FitSplineRegression}{$E_i$}
  \EndFor
  \\ \Return $\{f^c_i\}$
  
  \end{algorithmic}
\end{algorithm}

\noindent \textbf{\textit{Auto-tuner learning pseudocode.}}
We present the auto-tuner training pseudocode in Algorithm~\ref{algo:fit-auto-tuners}.
In line 1 to line 4, we traverse each inserted filter $M_i$, evaluate it on the calibration set $\mathcal{S}_{q(c)}$, calculate and sort the absolute prediction errors.
We then iterate over all sorted position in line 5.
For each sorted position $j$, we set the adjusting offset $o_{j_i}$ to be the sorted error $\alpha_{i, (j)}$ for each filter $M_i$ in line 7.
We simulate the LeaFi-enhanced search for the calibration queries $\mathcal{S}_{q(c)}$ and calculate their achieved result qualities $\{q_{j, k}\}$ in line 9 to line 10, and add the (adjust offset, result quality) pair $(q_{j, k}, o_{j, i})$ to $M_i$'s auto-tuner training set in line 12.
Finally, we train and return the auto-tunes $\{f^c_i\}$.

\subsubsection{Filter Prediction Auto-tuning in Search.}
\label{sec-conformal-search}
In LeaFi-enhanced search stage, given a user-requested quality target $q$, we first determine the corresponding adjusting offset $o_i$ using the learned mapping $f^{c}_i(q)$ for each learned filter $M_i$.
The filter prediction is adjusted accordingly, i.e., $d^{\text{F}}_{i}=d^{\text{f}}_{i}-o_i$, based on which the filter-based pruning for node $N_i$ is triggered.

\subsection{Complexity Analysis}
\label{sec-conformal-search}
In this section, we briefly analyze the time and space complexity for the LeaFi framework.
The cost of LeaFi is mainly from LeaFi-enhanced index building.
In LeaFi-enhanced search, the overhead of filter inference is expected to be covered by search time reduction through the leaf node selection procedure in Section~\ref{sec-method-node-selection}.

Among the four steps in LeaFi-enhanced index building, we note that the time bottleneck lies in training data generation and filter learning, rather than leaf node selection and conformal auto-tuner learning.
The complexity of training data generation is $\mathcal{O}(n_q \times n)$, linear to the training data size $n_q$ and dataset size $n$.
The filter training is proportional to the number of inserted filters $n_F$ as well as the training time %
of an individual filter $t_F$, resulting in a complexity of $\mathcal{O}(t_F \times n_F)$.
Hence, the time complexity of LeaFi enhancement in index building (i.e., besides the original index building time complexity) is $\mathcal{O}(n_{q}n + t_{F}n_{F})$.

The space complexity of LeaFi $\mathcal{O}(n_{q} + s_{F}n_{F})$, although proportional to the training data size $n_q$ and number of inserted filters $n_F$, is negligible compared to the dataset size $n$.

We note that with a reasonable training data size (2K series in our experiments) and a proper node size threshold ($th=a \cdot t^{\text{F}} / t^{\text{S}}$ in our experiments), LeaFi can provide considerable benefits when compared to the original indexes.

\section{Experiments}
\label{sec-experiments}

In this section, we report our experiment evaluation using two different tree-based series indexes and five diverse datasets.
The source code and datasets are available online\hyperref[fn:leafi-url]{\footnotemark[\value{leafi-url}]}.

\subsection{Evaluation Setup}
\label{sec-exp-setup}
The experiments were carried out on a server equipped with an Intel(R) Xeon(R) Gold 6242R CPU, 520 GB RAM, and an NVIDIA Quadro RTX 6000 with 24 GB GDDR6 memory. 
The software environments were gcc/9.4.0, cuda/11.2, libtorch/1.13.1 (for MLP), and gsl/2.7.2 (for spline regression).

\noindent \textbf{\textit{Datasets.}} 
We include a variety of datasets in our evaluation, consisting of one synthetic dataset and four real datasets from different domains.
For the synthetic dataset, we choose RandWalk~\cite{DBLP:conf/sigmod/FaloutsosRM94}, which is generated by accumulating steps following a standard Gaussian distribution $N(0,1)$. 
Regarding the real datasets, we select Seismic~\cite{SC:trabant2012data} from seismology, Astro~\cite{GS:soldi2014long} from astronomy, Deep~\cite{DBLP:conf/cvpr/YandexL16} and SIFT~\cite{DBLP:conf/icassp/JegouTDA11} from image processing.
Note that Deep and SIFT are two popular high-dimensional vector datasets of image descriptors (not data series).
The series length is 256, except for Deep (96) and SIFT (128).
Each dataset contains 25 million (i.e., 25M) data series in our experiments.
In addition, we experimented with datasets of sizes between 10M-100M to test the scalability of LeaFi.

Following recent data series studies~\cite{DBLP:journals/pvldb/EchihabiFZPB22}, we create four different query sets with varying levels of difficulty for every dataset.
Each query set consists of 1,000 series,  generated by adding 10\%, 20\%, 30\%, and 40\% Gaussian noise into uniform random samples.
Additionally, we generate one training set of 2,000 series for each dataset, by adding random levels of Gaussian noise $\in$ [10\%, 40\%].
The split of training and validation is 4:1.

\noindent \textbf{\textit{Tree-based series indexes.}} 
We choose MESSI~\cite{DBLP:conf/icde/PengFP20} (the SOTA variant of iSAX~\cite{DBLP:conf/kdd/ShiehK08}) and DSTree~\cite{DBLP:journals/pvldb/WangWPWH13} as the backbone indexes. %
The split threshold of node size is 10k for both indexes. 
We use 16 threads in MESSI for index building and query answering.

\noindent \textbf{\textit{LeaFi instantiation.}} 
We use MLP to instantiate learned filters. 
Each MLP model have one hidden layer, whose dimension is set the same as the input series. 
We train these models using the stochastic gradient descent (SGD)~\cite{GS:robbins1951stochastic} algorithm for at most 1,000 epochs.
The learning rate, initialized as 0.01, is divided by 10 (until $10^{-5}$) when the validation errors plateau.

We leverage multicore processing to accelerate the collecting of training data and the training of inserted filters.
For the efficient collecting of training data, we propose a two-pass strategy that works for all tree-based indexes.
In the first pass, we detach all the leaf nodes with inserted filters to calculate their node-wise nearest neighbor distances in parallel.
Then in the second pass, we search for the global queries and collect the best-so-far distances efficiently by reusing the distances calculated in the first pass.
For filter training, we assign one CUDA stream to each thread~\cite{web/nvidia/cuda} and use 16 threads to train the models in parallel.

\noindent \textbf{\textit{Comparison approaches.}} 
The exact search performance of the original indexes is an important baseline for LeaFi.
Additionally, we choose $\epsilon$-search, $\delta\epsilon$-search~\cite{DBLP:journals/pvldb/EchihabiZPB19}, ProS~\cite{DBLP:conf/sigmod/GogolouTEBP20, DBLP:journals/vldb/EchihabiTGBP23} and LT~\cite{DBLP:conf/sigmod/LiZAH20} to represent the early-stopping strategy, as well as LR~\cite{SC:kang2021case} to represent the reordering strategy.

Note that the comparison approaches do not natively support result quality targets. 
Moreover, LT employs sets of features designed specifically for graph-based indexes~\cite{DBLP:journals/pami/MalkovY20} and inverted indexes~\cite{DBLP:journals/pami/BabenkoL15}, which do not apply to the tree-based data-series indexes used in this study.
Hence, we need to adjust these approaches to fit our scenario, and then also pay the cost to fine-tune them. %
Specifically:
\begin{enumerate}[noitemsep, topsep=0pt, wide=\parindent]
  \item For $\epsilon$-search, we grid-search for the maximal $\epsilon \in [1, 7]$ that provides $\geq$ 99\% recall on the validation set~\cite{DBLP:journals/pvldb/EchihabiZPB19}.
  A larger $\epsilon$ results in larger acceleration but lower recall.
  We set $\epsilon$=1 when 99\% recall cannot be achieved.

  \item For $\delta\epsilon$-search, we set $\epsilon$=0 and then grid-search for the smallest $\delta \in [90\%, 99.9\%]$ that provides $\geq$ 99\% recall on the validation set~\cite{DBLP:journals/pvldb/EchihabiZPB19}.
  A larger $\delta$ results in smaller acceleration but higher recall.
  We set $\delta$=99.9\% when 99\% recall cannot be achieved.
  We estimate the nearest neighbor distances using the node-wise nearest neighbor distances on the validation set.
  
  \item For ProS, we choose the early-stopping strategy that predicts whether the nearest neighbor results have been found after checking certain number of leaf nodes using the best-so-far distances~\cite{DBLP:conf/sigmod/GogolouTEBP20, DBLP:journals/vldb/EchihabiTGBP23}.
  We set the early-stopping checking nodes to be [16, 64, 256, 512, 1024, 2048] for DSTree, and also [4096, 8192] for iSAX.

  \item For LT, we design our own features for tree-based series indexes following similar intuitions~\cite{DBLP:conf/sigmod/LiZAH20}: (a) the input query, (b) the ratio between the node-wise nearest neighbor distances of the first leaf node and the first [2, 4, 8, 16] leaf nodes for DSTree ([8, 16, 32, 64] for MESSI), (c) the best-so-far distances after examining the first [1, 2, 4, 8, 16] leaf nodes for DSTree ([1, 8, 16, 32, 64] for MESSI), and (d) the ratio between the node-wise nearest neighbor distance of the first leaf node and the best-so-far distance after examining the 16th node for DSTree (or 64th for MESSI).
  Besides the need for a new feature template, we also need to tune the multiplier, a hyperparameter that expands the predicted number of early-stop nodes.
  We grid-search for the minimal multiplier $\in [1, 20]$ that can provide $\geq$ 99\% recall on the validation set.
  A larger multiplier results in smaller acceleration but higher recall.
  We set multiplier=20 when 99\% recall cannot be achieved. 

  \item For LR~\cite{SC:kang2021case}, we use the optimal reordering, i.e., the nearest neighbor results lie in the first examined leaf node, to provide the largest possible acceleration by any reordering strategies.
\end{enumerate}

\noindent \textbf{\textit{Evaluation measures.}} 
\label{sec-exp-measure}
We report the query time (lower is better), recall-at-1~\cite{DBLP:conf/sigmod/LiZAH20} (higher is better) and series pruning ratio (higher is better) in our experiments.

\subsection{Main Results}
\label{sec-exp-main-result}
\begin{figure}[tb]
  \centering
  
  \subfloat{
    \hspace{-.135\linewidth}
    \includegraphics[width=0.85\linewidth]{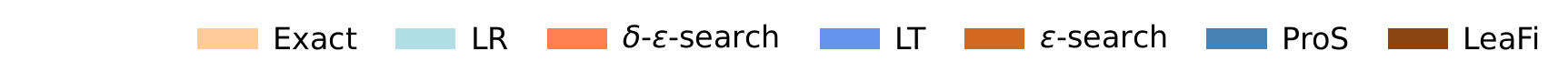}}\\[-2.5ex]
  \setcounter{subfigure}{0}
  
  \hspace{-.045\linewidth}
  \subfloat[Query time (DSTree)
    \label{fig:exp-main-time-dstree}]{
    \includegraphics[width=.33\linewidth]{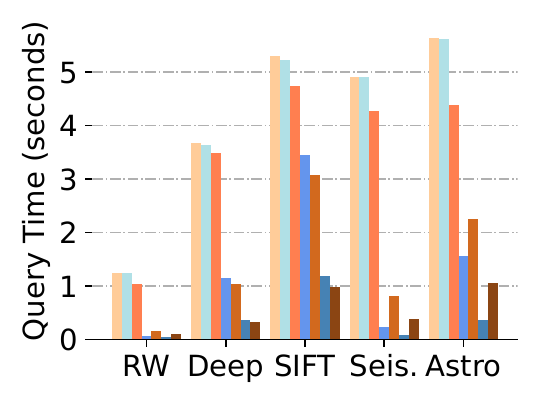}}
  \subfloat[Recall for 99\% target (DSTree)
    \label{fig:exp-main-recall-dstree}]{
    \includegraphics[width=.33\linewidth]{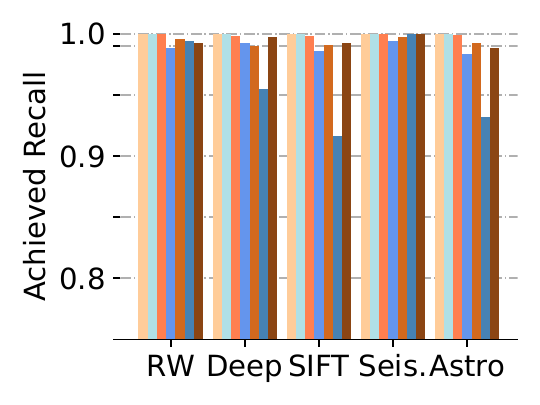}} \\
    
  \hspace{-.045\linewidth}
  \subfloat[Query time (MESSI)
    \label{fig:exp-main-time-isax}]{
    \includegraphics[width=.33\linewidth]{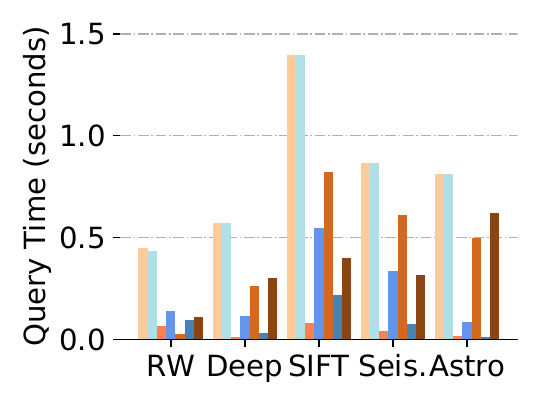}}
  \subfloat[Recall for 99\% target (MESSI)
    \label{fig:exp-main-recall-isax}]{
    \includegraphics[width=.33\linewidth]{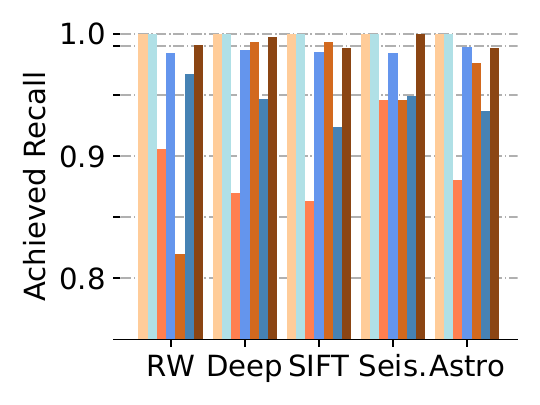}} \\
    
  \caption{
    The query time and achieved recalls at a target 99\% recall for enhanced DSTree and MESSI indexes across datasets.
  }
  
  \label{fig:exp-main-time-recall}
\end{figure}

\begin{figure*}[tb]
  \centering
  \subfloat{
  \includegraphics[width=0.99\textwidth]{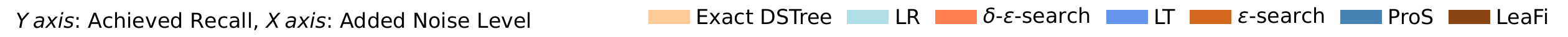}}\\[-3ex]
  \setcounter{subfigure}{0}
  
  \subfloat[RandWalk
    \label{fig:exp-detail-a}]{
    \includegraphics[width=.195\textwidth]{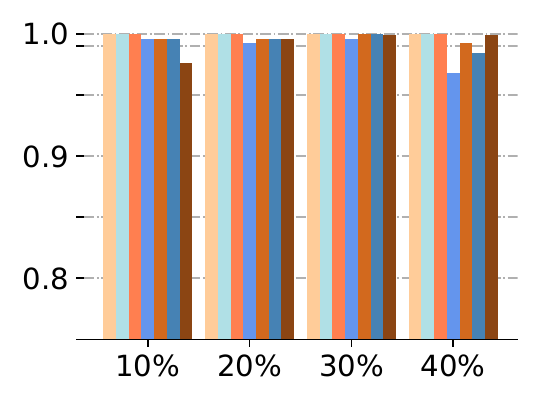}}
  \subfloat[Deep]{
    \includegraphics[width=.195\textwidth]{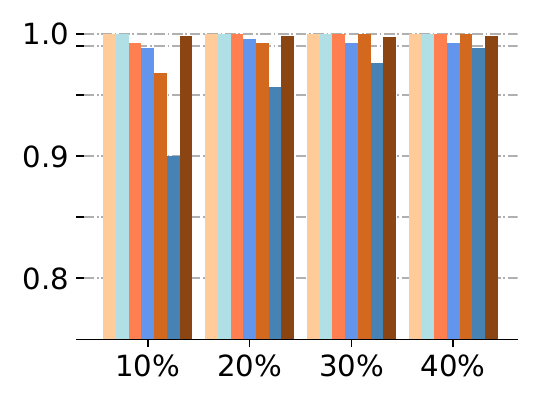}}
  \subfloat[SIFT]{
    \includegraphics[width=.195\textwidth]{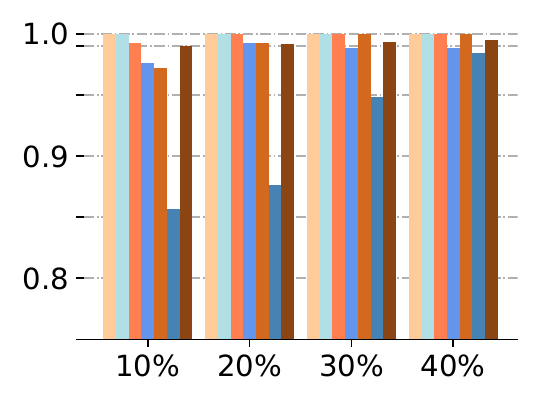}}
  \subfloat[Seismic]{
    \includegraphics[width=.195\textwidth]{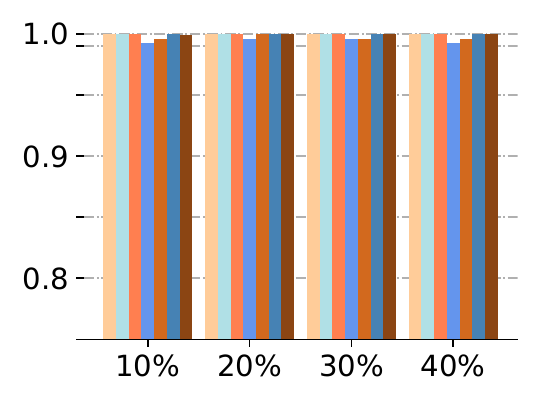}}
  \subfloat[Astro
    \label{fig:exp-detail-e}]{
    \includegraphics[width=.195\textwidth]{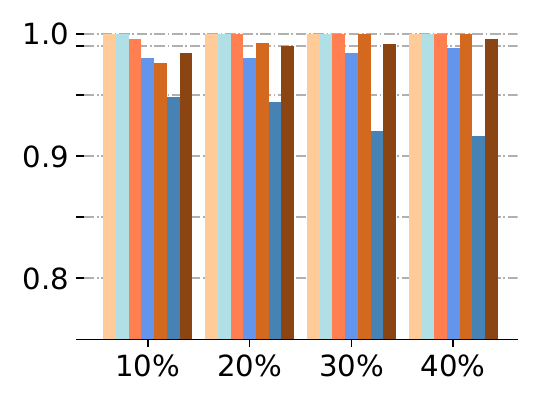}}\\
  
  \subfloat{
  \includegraphics[width=0.99\textwidth]{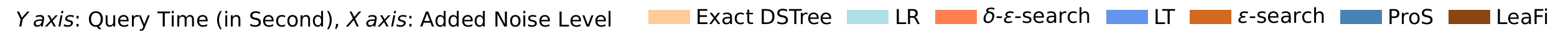}}\\[-3ex]
  \setcounter{subfigure}{5}
  
  \subfloat[RandWalk
    \label{fig:exp-detail-f}]{
    \includegraphics[width=.195\textwidth]{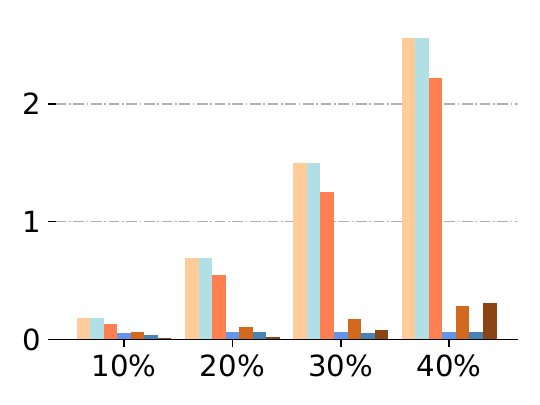}}
  \subfloat[Deep]{
    \includegraphics[width=.195\textwidth]{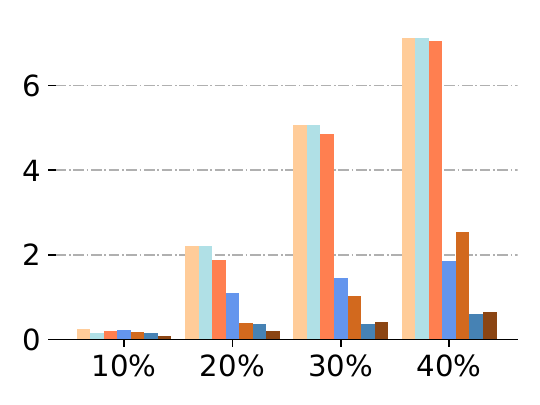}}
  \subfloat[SIFT]{
    \includegraphics[width=.195\textwidth]{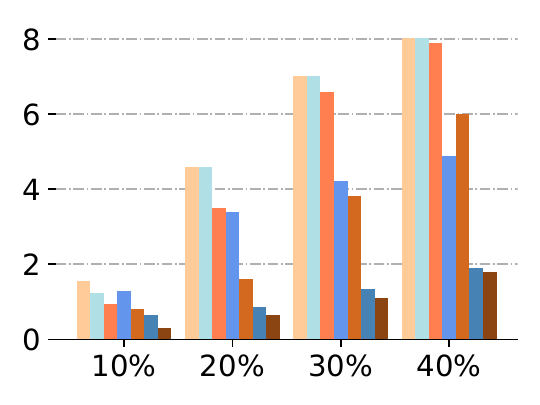}}
  \subfloat[Seismic]{
    \includegraphics[width=.195\textwidth]{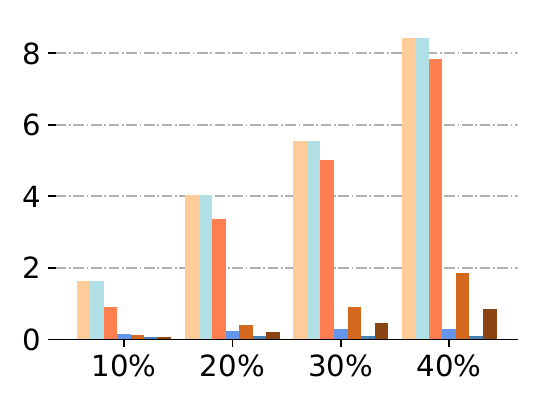}}
  \subfloat[Astro
    \label{fig:exp-detail-j}]{
    \includegraphics[width=.195\textwidth]{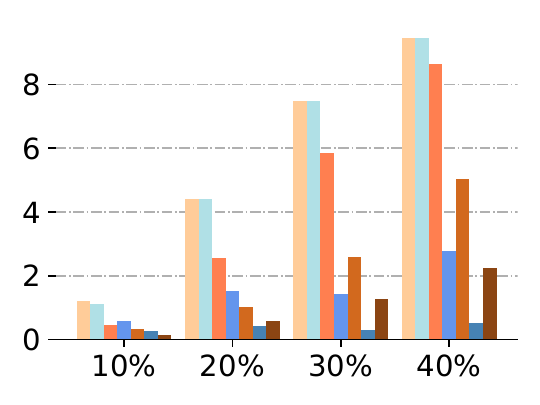}} \\
  
  \subfloat{
  \includegraphics[width=0.99\textwidth]{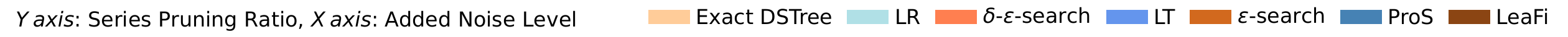}}\\[-3ex]
  \setcounter{subfigure}{10}
  
  \subfloat[RandWalk
    \label{fig:exp-detail-k}]{
    \includegraphics[width=.195\textwidth]{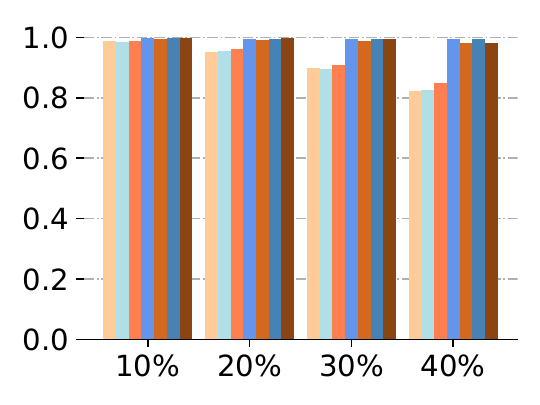}}
  \subfloat[Deep]{
    \includegraphics[width=.195\textwidth]{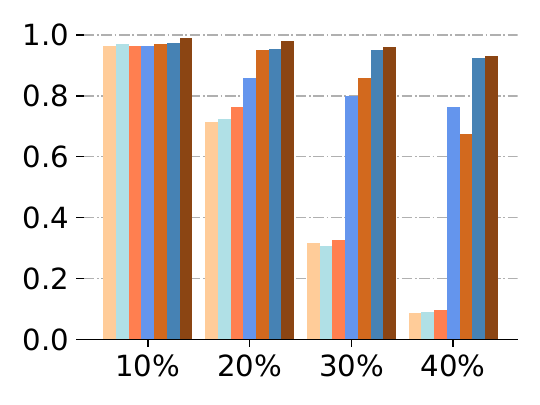}}
  \subfloat[SIFT]{
    \includegraphics[width=.195\textwidth]{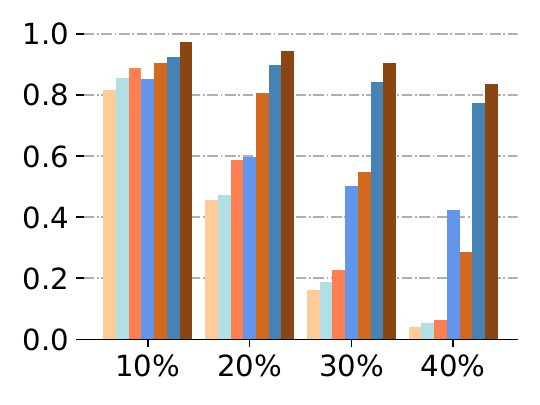}}
  \subfloat[Seismic]{
    \includegraphics[width=.195\textwidth]{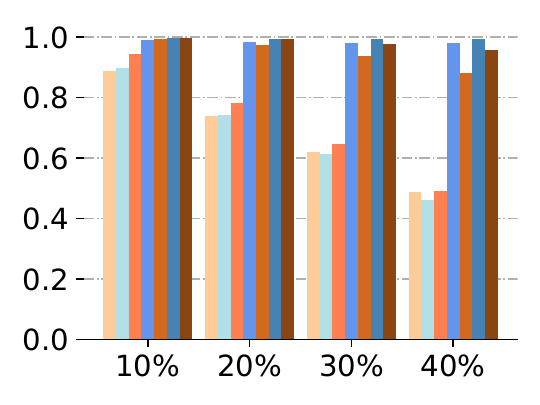}}
  \subfloat[Astro
    \label{fig:exp-detail-o}]{
    \includegraphics[width=.195\textwidth]{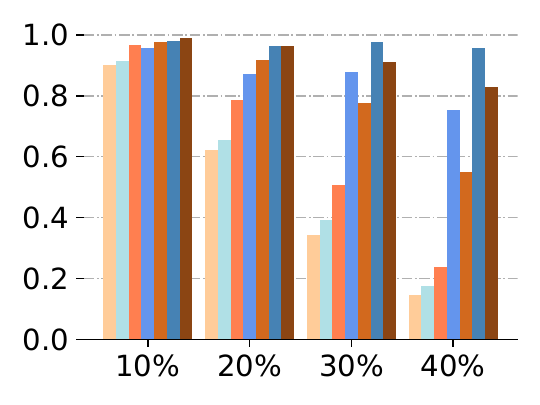}}
  
  \caption{
    The actual recall (\ref{fig:exp-detail-a} to \ref{fig:exp-detail-e}), average query time (\ref{fig:exp-detail-f} to \ref{fig:exp-detail-j}) and average series pruning ratio (\ref{fig:exp-detail-k} to \ref{fig:exp-detail-o}) for enhanced DSTree indexes, targeting at 99\% recall.
  }
  \label{fig:exp-detail-dstree-recall-time-prune}
  
\end{figure*}

We first report the average query time and actual recalls for a recall target 99\% in Figure~\ref{fig:exp-main-time-recall}.
We also provide a detailed analysis for the enhanced DSTree indexes by reporting the detailed results over different query noise levels, in Figure~\ref{fig:exp-detail-dstree-recall-time-prune}.

Overall, we find that LeaFi is the only solution that can provide a substantial search time improvement while achieving 99\% recall in all test cases.
LeaFi-enhanced series indexes improved pruning ratio by up to 20x (by DSTree on SIFT dataset with queries of 40\% noise) and search time by up to 32x (by DSTree on RandWalk dataset with queries of 20\% noise), while maintaining a target recall of 99\%. 
Moreover, LeaFi is the only solution that supports ad-hoc quality targets, chosen at query time, independently for each query.

On the contrary, despite our best efforts to tune the comparison methods, none of the early-stopping strategies ($\delta\epsilon$-search, LT, $\epsilon$-search, and ProS) can consistently achieve 99\% recall %
across four query subsets of different noise levels. %
The reordering strategy (LR) failed to provide query time improvement.
These empirical observations comply to our analysis in Section~\ref{sec-lit-learned-index}.

\subsubsection{Main Results on DSTree}
\label{sec-exp-main-dstree}
We first discuss the average search time and actual recalls for a recall target 99\% on DSTree in Figure~\ref{fig:exp-main-time-dstree} and \ref{fig:exp-main-recall-dstree}.
The split-down results across different query noise levels are reported in Figure~\ref{fig:exp-detail-dstree-recall-time-prune}.

LeaFi-enhanced DSTree obtained 99\% recall on all 5 datasets, with an average speedup of 9.2x, and best speedup of 12.7x, when compared to the exact search.
This is attributed to the largely improved pruning capability of DSTree leaf nodes.
LeaFi achieved an average pruning ratio of 95.6\%, compared to 59.8\% without any learned filters (cf. Figure~\ref{fig:exp-detail-dstree-recall-time-prune}).
$\epsilon$-search and $\delta\epsilon$-search for DSTree indexes also provided 99\% recall on 5 datasets.
However, LeaFi-enhanced DSTree are much more efficient, i.e., 2.4x faster than $\epsilon$-search and 8.1x faster than $\delta\epsilon$-search on average.
This observation indicates that although heuristic early-stopping approaches can effectively support quality targets, machine learning-based enhancements can provide larger query acceleration.
We also observe that $\delta\epsilon$-search requires the accurate estimation of the nearest neighbor distances~\cite{DBLP:journals/pvldb/EchihabiZPB19}, which is extremely hard.

ProS- and LT-enhanced DSTree achieved 99\% recall on only 2 datasets, which are RandWalk and Seismic for ProS, and Deep and Seismic for LT.
The average recall of ProS is 96\%, largely lagging behind the 99\% target.
Although the average recall of LT is 98.9\%, LeaFi-enhanced DSTree are 1.9x faster than LT on average.
LT's inability to achieve 99\% recall by tuning the multiplier also implies that the model training for LT is much harder than LeaFi.
This is because LeaFi fits one model to one leaf node with a small target range $\in$[0, 10], while LT fits one model to the whole datasets with large target range [$10^2$, $10^5$].

\noindent \textbf{\textit{Detailed analysis at different query noise levels.}} 
Breaking down the results across different query noise levels, we report the average search time, actual recalls and pruning ratios in Figure~\ref{fig:exp-detail-dstree-recall-time-prune}.
LeaFi-enhanced DSTree achieved 99\% recall on 17 of 20 (85\%) cases, with an average of 99.4\% and the lowest of 97.6\%.
The 3 cases of <99\% recall are with queries of small noise levels, 10\% on RandWalk, 10\% and 20\% on Astro.
This is because these cases are equipped with much smaller nearest neighbor distances than cases of higher noise, hence challenging LeaFi's conformal auto-tuners.
The best speedup of LeaFi-enhanced DSTree is 32.4x, obtained on RandWalk with queries of 20\% noise.
LeaFi achieved an average pruning ratio of 95.6\%, compared to 59.8\% without any learned filters.

Similarly to the aggregated results in Figure~\ref{fig:exp-detail-dstree-recall-time-prune}, $\epsilon$-search and $\delta\epsilon$-search perform well in achieving the 99\% recall target, but with small time performance improvements.
$\epsilon$-search achieved 99\% recall on all 20 cases, while $\delta\epsilon$-search achieved 99\% recall on 17 of 20 (85\%) cases.
On the other hand, ProS- and LT-enhanced DSTree struggled in achieving 99\% recall target.
ProS-enhanced DSTree achieved 99\% recall on 7 (35\%) cases, while LT-enhanced DSTree achieved 99\% recall on 11 (55\%) cases.
LR, as expected, provided 100\% recall with marginal improvement in terms of search time and pruning ratio, compared to LeaFi.

In conclusion, LeaFi is the only solution that can provide a substantial improvement in search time (up to 32.4x faster) and pruning ratio (up to 20x more), while consistently maintaining 99\% recall. %

\subsubsection{Main Results on iSAX}
\label{sec-exp-main-isax}
We provide the average search time and actual recalls for a recall target 99\% on MESSI in Figure~\ref{fig:exp-main-time-isax} and \ref{fig:exp-main-recall-isax}.
Due to the lack of space, we remove the splitting-down figures across query noise levels to an extended version.

MESSI differs from DSTree in the following aspects: (1) MESSI has more leaf nodes, yet with smaller filling factors, than DSTree; (2) MESSI can provide tighter node summarization than DSTree, while DSTree better groups similar series into the same leaf nodes~\cite{DBLP:journals/pvldb/AziziEP23}.
Moreover, the fact that MESSI utilizes the parallelization capabilities of modern hardware is an additional factor that affects performance. 
These elements explain the performance differences between enhanced DSTree and MESSI.

LeaFi improved MESSI query time on all 5 datasets while maintaining 99\% recall, with an average speedup of 2.7x and the best speedup of 4x.
In the splitting-down results across query noise levels, the best speedup is 13x on RandWalk with queries of 20\% noise.
The search time improvement is less than the pruning ratio improvement, which is improved from 55.2\% to 86.2\% on average (which is expected to provide a 3.2x speedup).
We believe this is due to GPU resource contention as a result of concurrent filter inference requests. %
We further verified this speculation by varying the leaf node thresholds in Section~\ref{sec-exp-detailed-node-size-thresholds}.
These results verified LeaFi can provide a substantial improvement in search time and pruning ratio, while achieving 99\% recall for MESSI in a multithread environment.

For the comparison methods, only LT-enhanced MESSI achieved 99\% recall on only 2 of the datasets.
Other comparison methods cannot achieve 99\% recall in any dataset, despite thorough tuning using the validation set.
In the detailed results across query noise levels, LeaFi-enhanced MESSI achieved 99\% recall on 14 of 20 (70\%) cases, while $\epsilon$-search and LT on 6 (30\%) cases, $\delta\epsilon$-search and ProS on 1 (5\%) case.
Moreover, LeaFi-enhanced MESSI is  faster than LT-enhanced MESSI on 3 datasets. 

In summary, the results on MESSI are consistent with those on DSTree. %
LeaFi is the only solution that maintains 99\% recall %
across queries of all noise levels, %
while providing considerable query speedups (up to 13x) and pruning ratio improvements (up to 76.2\%).

\subsection{Extended Analysis}
\label{sec-exp-detailed-result}
\begin{figure}[tb]
  \centering
  
  \subfloat{
    \hspace{-.14\linewidth}
    \includegraphics[width=0.85\linewidth]{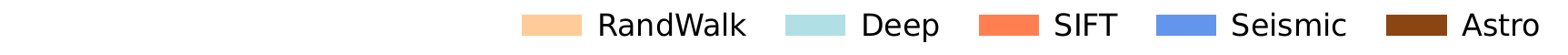}}\\[-3.2ex]
  \setcounter{subfigure}{0}
  
  \hspace{-.02\linewidth}
  \subfloat[DSTree
    \label{fig:exp-recalls-dstree}]{
    \includegraphics[width=.33\linewidth]{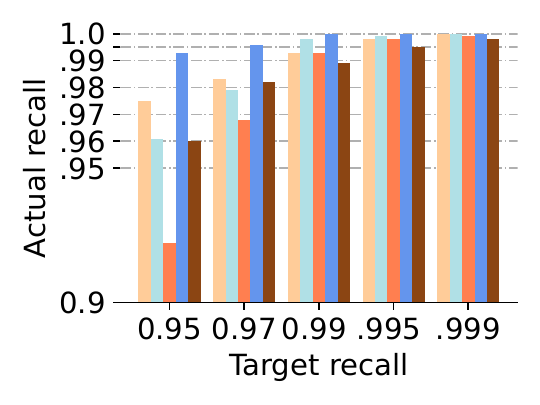}}
  \hspace{.05\linewidth}
  \subfloat[MESSI
    \label{fig:exp-recalls-isax}]{
    \includegraphics[width=.33\linewidth]{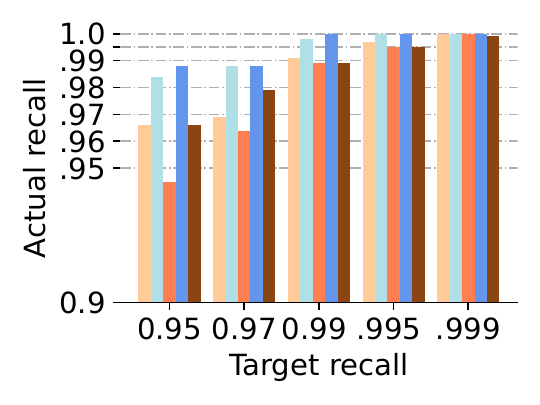}}
    
  \caption{
    The target recall and actual (achieved) recall for LeaFi-enhanced DSTree and MESSI indexes across datasets. 
  }
  
  \label{fig:exp-combined-recalls}
\end{figure}

\begin{figure}[tb]
  \centering
    \subfloat{
    \hspace{-.23\linewidth}
    \includegraphics[width=0.85\linewidth]{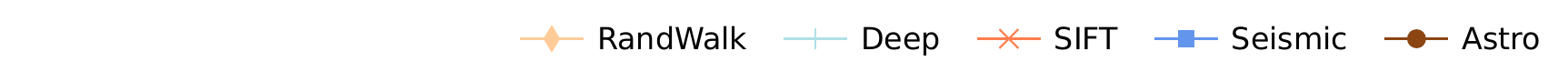}}\\[-3.2ex]
  \setcounter{subfigure}{0}
  
  \subfloat[DSTree
    \label{fig:exp-scale-recall-dstree}]{
    \includegraphics[width=.3\linewidth]{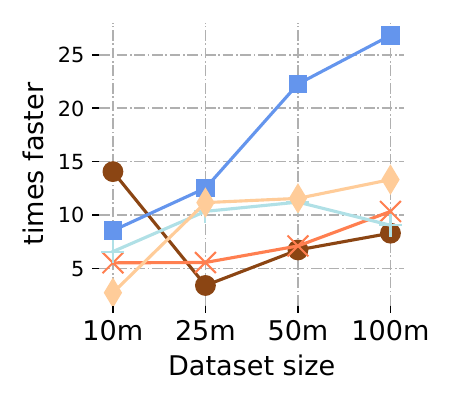}}
  \hspace{.05\linewidth}
  \subfloat[MESSI
    \label{fig:exp-scale-recall-isax}]{
    \includegraphics[width=.3\linewidth]{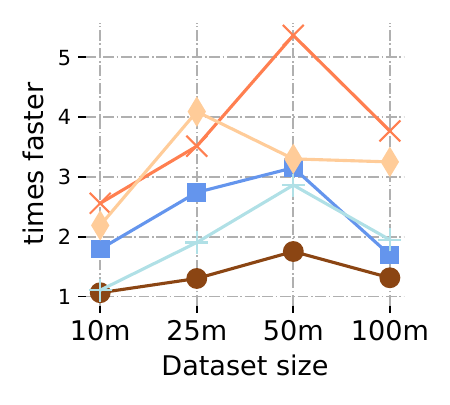}}
    
  \vspace*{-0.2cm}
  \caption{
    The query time improvements at a target 99\% recall for LeaFi-enhanced DSTree and MESSI when varying the dataset size between 10M-100M.
  }
  
  \label{fig:exp-scale-time-recall}
\end{figure}

\begin{figure}[tb]
  \centering
  
  \subfloat[Query time vs. node size threshold
    \label{fig:exp-node-threshold-time}]{
    \includegraphics[width=.28\linewidth]{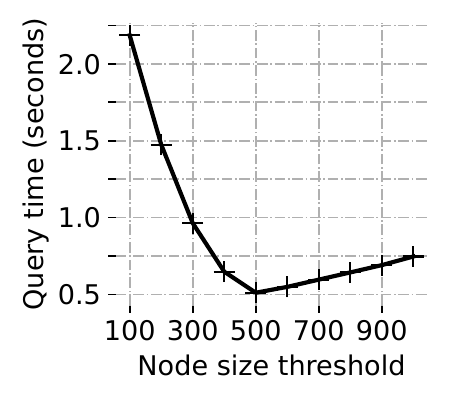}} 
  \hspace{.06\linewidth}
  \subfloat[Number of inserted filters vs. node size threshold
    \label{fig:exp-node-threshold-number-filters}]{
    \includegraphics[width=.28\linewidth]{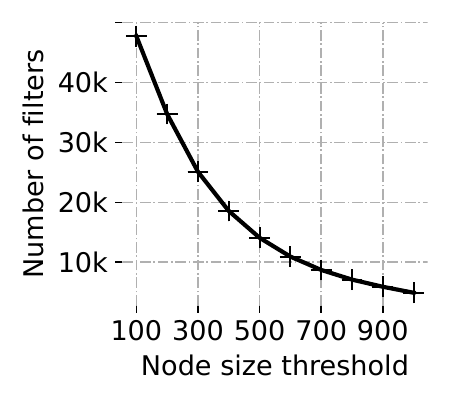}} \\
    
  \subfloat[Pruning ratio vs. node size threshold
    \label{fig:exp-node-threshold-prune}]{
    \includegraphics[width=.28\linewidth]{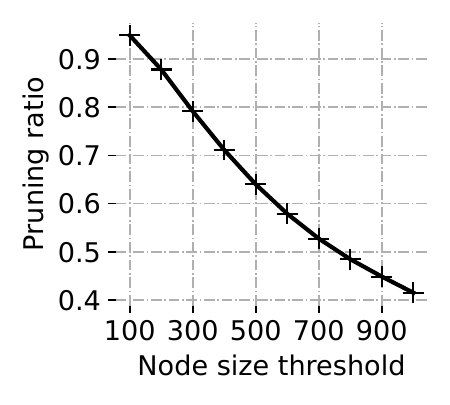}}
  \hspace{.06\linewidth}
  \subfloat[%
    Query time vs. GPU memory size%
    \label{fig:exp-mem-time}]{
    \includegraphics[width=.28\linewidth]{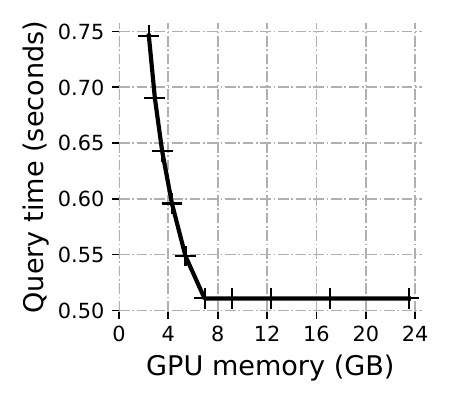}}
    
  \caption{
  Tradeoff between the available GPU memory and leaf node size threshold $th$, and the query time, pruning ratio, and number of inserted filters (for LeaFi-enhanced MESSI).%
  }
  
  \label{fig:exp-node-threshold}
\end{figure}

In this section, we report the actual recalls under different recall targets in Figure~\ref{fig:exp-combined-recalls}, the query time for 100M datasets in Figure~\ref{fig:exp-scale-time-recall}, and the query time across different leaf node threshold in Figure~\ref{fig:exp-node-threshold} to better understand the performance of LeaFi-enhanced indexes.

\subsubsection{Varying Recall Target}
\label{sec-exp-detailed-recall}
We present the actual recalls of LeaFi-enhanced DSTree and MESSI for user-requested high recalls $\in$[95\%, 99.9\%] in Figure~\ref{fig:exp-combined-recalls}.
These results verify that the conformal auto-tuner of LeaFi can effectively adjust the filter predictions to accommodate user quality targets.  

Overall, LeaFi-enhanced DSTree achieved the recall targets on 31 of 35 (88.6\%) cases and LeaFi-enhanced MESSI achieved the recall targets on 28 of 35 (80\%).
The recall difference when the target recall cannot be achieved is marginal, i.e., 0.54\% for DSTree and 0.35\% for MESSI.
The largest difference was found by DSTree on the SIFT dataset, with a 95\% recall target.
We believe that the small number of validation examples (300) resulted in imbalanced statistics, leaving some queries sensitive to the adjusting offset for the 95\% recall target.
Even though our focus has been on very high recall targets, we note that LeaFi-enhanced indexes are effective for smaller recall targets, as well.
As the recall target decreases, LeaFi leads to increased speedups.
We omit these results for brevity.

\subsubsection{Varying Dataset Size}
\label{sec-exp-larger}
In Figure~\ref{fig:exp-scale-time-recall}, we present the query time improvements of LeaFi-enhanced DSTree and MESSI for 99\% recall target as we vary the dataset size between 10M-100M.
We note that in this experiment, we build a different index structure for each dataset size (i.e., a tree with a different number of leaves and different sets of series in each leaf), which leads to a different set of inserted and trained filters in each case. 
This explains the non-smooth curves shown in the graphs.

Overall, the results demonstrate that LeaFi-enhanced series indexes 
consistently improve query answering times across all dataset sizes (while maintaining the 99\% target recall). 
They also show a trend for larger query time improvements as the dataset size grows larger.
Delving into the results, we observe that LeaFi-enhanced DSTree obtained an average speedup of 9.2x on the 25M datasets, and a 13.6x average speedup on the 100M datasets.
This increase in speedup is attributed to the increased pruning ratio for the 100M dataset (from 94.5\% to 97.6\%).
On the other hand, LeaFi-enhanced MESSI obtained a 2.7x average speedup on the 25M datasets, %
and a 2.4x speedup on the 100M datasets.
This behavior is due to the increased number of inserted filters (from $\sim$10k to $\sim$30k), which overwhelms the GPU access.
Exploring techniques for efficient GPU utilization %
would be a promising direction going forward~\cite{DBLP:conf/mlsys/0002YZLP21}.

\subsubsection{Varying Node Size Threshold}
\label{sec-exp-detailed-node-size-thresholds}
Figures~\ref{fig:exp-node-threshold}a-c depict the search time and pruning ratios of LeaFi-enhanced MESSI across a range of leaf node thresholds on the 25M Deep dataset, with queries of 40\% noise. %
With the increase of leaf node size threshold $th$, the search time first decreased then increased, while the pruning ratios kept decreasing.
The decrease of the search time came from both the filter inference time and the multithreading GPU access competition overhead.
Empirically, we measured $t^{\text{F}}/t^{\text{S}} \approx 279$ for the Deep dataset.
By setting the hyperparameter $a=2$, LeaFi used $th=558$ for Deep dataset, which provided near-optimal performance on iSAX. 
Without the multithreading GPU access competition, LeaFi-enhanced DSTree has a turning point of $\sim$100.

\subsubsection{Varying GPU Memory}
\label{sec-exp-detailed-gpu-memory}
We present in Figure~\ref{fig:exp-mem-time} the search time of LeaFi-enhanced MESSI across a range of GPU memory limits under the same setting as in Figures~\ref{fig:exp-node-threshold}a-c.
All reported results targeted at 99\% recall, and achieved at least 99.3\% actual recall.
Overall, the query time starts to decrease and then stabilizes as we increase the available GPU memory.
The decreasing phase is bounded by GPU memory, while the stabilization phase is bounded by the number of filters that are expected to improve query time.
The point where improvement stops is at 6.9 GB, which corresponds to inserting filters to the 14K leaf nodes with more than $th=500$ series.
Inserting more filters would mean that these additional filters would end up in leaf nodes with less the $500$ series: in such cases, the filter inference time would be larger than the time to scan all series in the leaf, bringing no additional benefit. 
(The LeaFi-enhanced DSTree results are similar, and omitted for brevity.) %

\begin{figure}[tb]
  \centering
  \begin{minipage}[c]{0.33\linewidth}
    \centering
    \captionof{table}{%
    Filter inference time ($\mu$s) and the node size threshold $th$ derived for Deep (length 96).
    }
    \label{tab:exp-cnn}
    \small
    \begin{tabular}{c|r|r}
    \toprule
      Type & Time & $th$ \\ %
    \midrule
      MLP & 46 & 558\\ \hline
      CNN & 258 & 3,129 \\ \hline
      RNN & 11,695 & 141k \\ %
    \midrule
      $d(\cdot,\cdot)$ & 0.16 & - \\
    \bottomrule
    \end{tabular}
  \end{minipage}
  \hspace{0.04\linewidth}
  \begin{minipage}[c]{0.33\linewidth}
    \centering
    \includegraphics[width=\linewidth]{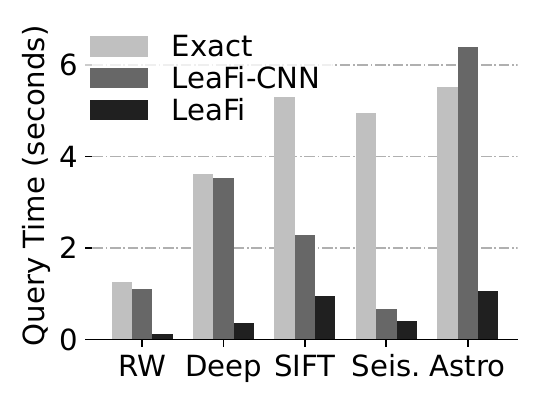} 
    \caption{%
    LeaFi with CNN filters on DSTree.
    }%
    \label{fig:exp-cnn}
  \end{minipage}
\end{figure}

\subsubsection{Varying Model Type}
\label{sec-exp-detailed-model-type}
Table~\ref{tab:exp-cnn} and Figure~\ref{fig:exp-cnn} illustrate the results on instantiating LeaFi filters with Convolutional Neural Network (CNN) and Recurrent Neural Network (RNN) models.
The CNN model has 2 convolutional layers with kernel size 3 and latent channel size equal to the series length.
Similarly, the RNN model contains 2 LSTM blocks.
Both CNN and RNN are much more computationally expensive than MLP, resulting in 5.6x and 254x larger node size thresholds.
These high thresholds render the deployment of RNN filters not suitable in our setting (the RNN inference time would only be justified for leaf nodes of size $\geq 11k/0.16*2=141K$ series, while our leaves have a maximum capacity of 10K).
Having $th=3k$, the benefit of CNN is also limited, leading to a much smaller number of inserted filters (when compared to MLP), and in general, significantly smaller pruning ratios and query time improvements.

\begin{table}[tb]
  \caption{%
  Query time (seconds) and actual recall for LeaFi-enhanced DSTree without ($-$Local) and with ($+$Local) local training data, for 99\% recall target; $+$Local corresponds to our proposed LeaFi solution.
  }%
  \label{tab:exp-nonlocal}
{\small
  \begin{tabular}{@{}c|ccc|lr@{}}
    \toprule
    \multicolumn{1}{c}{ } & \multicolumn{3}{c}{Query time (s)} & \multicolumn{2}{c}{Actual recall}\\ 
    \cmidrule(rl){2-4} \cmidrule(rl){5-6}
    Dataset & Exact & $-$Local & $+$Local & $-$Local & $+$Local \text{ } \\
    \midrule
    RandWalk & 1.2 & 0.4 & \textbf{0.1} & \text{ } 0.98 \ding{55} & 1.0 \ding{51} \text{ } \\
    Deep & 3.6 & 0.7 & \textbf{0.3} & \text{ } 0.96 \ding{55} & 0.99 \ding{51} \text{ } \\
    SIFT & 5.3 & 1.3 & \textbf{1.0} & \text{ } 0.94 \ding{55} & 0.99 \ding{51} \text{ } \\
    Seismic & 4.9 & \textbf{0.3} & 0.4 & \text{ } 0.99 \ding{51} & 1.0 \ding{51} \text{ } \\
    Astro & 5.5 & \textbf{1.0} & 1.6 & \text{ } 0.91 \ding{55} & 0.99 \ding{51} \text{ } \\
    \bottomrule
  \end{tabular}
}
\end{table}

\subsubsection{Without Local Training Data}
\label{sec-exp-detailed-no-local}
We study the impact of local training data in Table~\ref{tab:exp-nonlocal}.
We observe that training LeaFi without local training data can still bring improvements in query time, but it cannot consistently achieve the 99\% recall target.
This is because the nodewise NN distances of the global queries are larger than the distances of their query results.
Removing the local training queries also removes the query results' distance ranges from the training data, making these lower ranges ignored by the filters and conformal auto-tuners.
These results confirm that local training data is necessary for LeaFi.

\begin{figure}[tb]
  \centering
  \begin{minipage}[t]{0.48\linewidth}
    \centering
    \captionof{table}{%
    Indexing time breakdown (minutes) for LeaFi-enhanced DSTree and MESSI on Seismic 100M.
    }%
    \label{tab:exp-training-time}
    \small
  \begin{tabular}{@{}c|c|c@{}}
    \toprule
     & DSTree & MESSI \\
    \midrule
    Indexing & 28.6 & 5.3 \\
    Collecting data & 104.2 & 67.5 \\
    Training & 46.4 & 48.3 \\
    \bottomrule
  \end{tabular}
  \end{minipage}
  \hspace{0.04\linewidth}
  \begin{minipage}[t]{0.46\linewidth}
    \centering
    \captionof{table}{%
    Additional space overhead (GB) for LeaFi-enhanced DSTree and MESSI on Seismic 100M.
    }%
    \label{tab:exp-space-overhead}
    \small
  \begin{tabular}{@{}c|c|c@{}}
    \toprule
     & DSTree & MESSI \\
    \midrule
    Data & 100 & 100 \\
    Index structure & 1.5 & 0.3 \\
    Filters & 4.7 & 4.8 \\
    \bottomrule
  \end{tabular}
  \end{minipage}
\end{figure}

\subsubsection{Training Time}
\label{sec-exp-detailed-train-time}
We report the indexing and training time for LeaFi-enhanced indexes in Table~\ref{tab:exp-training-time}.
LeaFi-enhanced DSTree spends more time in indexing and collecting training data, because it employs a more complex summarization (EAPCA), while LeaFi-enhanced MESSI needs longer training time, because it employes more filters than DSTree (19k vs. 18k).
We note that if training time is a critical resource, the users can choose to train fewer filters, yet, still benefit from the improved pruning ratios.
As Figure~\ref{fig:exp-node-threshold}a shows, reducing the number of filters from $\sim$14k to $\sim$7k reduces the training time to half, while the query time only increases from $\sim$0.5 to $\sim$0.75 seconds, still outperforming the baseline.

\subsubsection{Space Overhead}
\label{sec-exp-detailed-space}
We report the space overhead for LeaFi-enhanced indexes in Table~\ref{tab:exp-training-time}.
The additional space needed by the filters is comparable for DSTree and MESSI, and is in both cases a small percentage ($<$5\%) of the data size.

\section{Conclusions and Future Work}
\label{sec-conclusions}
In this paper, we present LeaFi, a framework that enhances tree-based series indexes with learned filters in order to accelerate data series similarity search, while satisfying a user-defined target recall.
LeaFi can improve pruning ratio by up to 20x, %
and query answering time by up to 32x, while maintaining a target recall of 99\%. %
These results set the foundations for future advancements in employing learned filters for data series similarity search,  
including the development of algorithms for inserting filters into both internal nodes and leaf nodes, estimating filter-based pruning ratios effectively, choosing among different learned filter models, %
and supporting updates. %
As updates may trigger node splitting, incremental learning~\cite{DBLP:journals/natmi/VenTT22} could play an important role in efficiently training the new filters of the children nodes based on the filter of the parent~\cite{DBLP:journals/pvldb/ChenGCT23}.
It would also be interesting to study the potential of LeaFi to enhance the performance of other index types, such as inverted indexes~\cite{DBLP:journals/pami/BabenkoL15} %
and Locality-Sensitive Hashing (LSH)~\cite{DBLP:journals/pvldb/HuangFZFN15}.%

\begin{acks}
Work partially supported by EU Horizon projects AI4Europe (101070000), TwinODIS (101160009), ARMADA (101168951), DataGEMS (101188416) and RECITALS (101168490).
\end{acks}

\bibliographystyle{ACM-Reference-Format}
\bibliography{main}

\end{document}